\documentclass{iopart}
\usepackage{graphicx}
\usepackage{iopams}
\usepackage{color}
\pdfoutput=1
\begin{document}
\title[Large morphological sensitivity of the magnetothermopower 
       in Co/Cu multilayers]
      {Large morphological sensitivity of the magnetothermopower 
       in Co/Cu multilayered systems}
\author{Voicu Popescu and Peter Kratzer}
\address{Faculty of Physics and
Center for Nanointegration (CENIDE),
University of Duisburg-Essen,
Lotharstra{\ss}e\ 1, 47057 Duisburg, Germany}
\ead{voicu.popescu@uni-due.de}
%
\begin{abstract}
We present results of first-principles calculations 
on the transport properties, both under an electric field
or a temperature gradient, in the Co/Cu  multilayered systems. 
The various effects brought about by the changes in the 
morphological parameters, such as the 
number of repeats and the layer thickness,
are discussed in a systematic way. 
Our calculations show that the Seebeck coefficient
and the magnetothermopower (MTP) converge
rather rapidly with the number of Co repeats.
In the range of thin Co layers, we find strong variations in 
amplitude and sign of both the Seebeck coefficient and the MTP.
These large variations, which have no correspondent 
in the (magneto)conductance, are shown to be the result of
quantum well states present in the minority spin channel of thin Co
layers.
\end{abstract}

\section{Introduction}

Metallic heterostructures of alternating magnetic and non-magnetic
materials have been in the focus of research for more than
two decades. These intense experimental and theoretical investigations
have been triggered by the giant magnetoresistance (GMR) effect
\cite{GB97}. Currently the key-stone of standard magnetic
field sensors, the GMR denotes the large change in the 
resistance caused by the switching
from an anti-parallel to a parallel
magnetic alignment of the adjacent magnetic layers 
under an external magnetic field.
An analogous phenomenon could be observed in multilayered 
structures subject to a {\em temperature gradient}, in which case
the central quantity measuring the magnetic response
was the magneto-thermopower (MTP). These experiments, 
performed both in the current-in-plane (CIP) 
\cite{CBM+91,PFS+92,SYPS93,NSH+94} as well as in the
current-perpendicular-to-the-plane (CPP) 
geometry \cite{BSO00}, mark the first successful 
attempts of linking the heat flow with the spin degree
of freedom, paving the way towards the 
emerging field of spin calorics \cite{SMAZF12}.

\begin{figure*}[t]
  \centering
  \includegraphics[width=0.8133\textwidth]{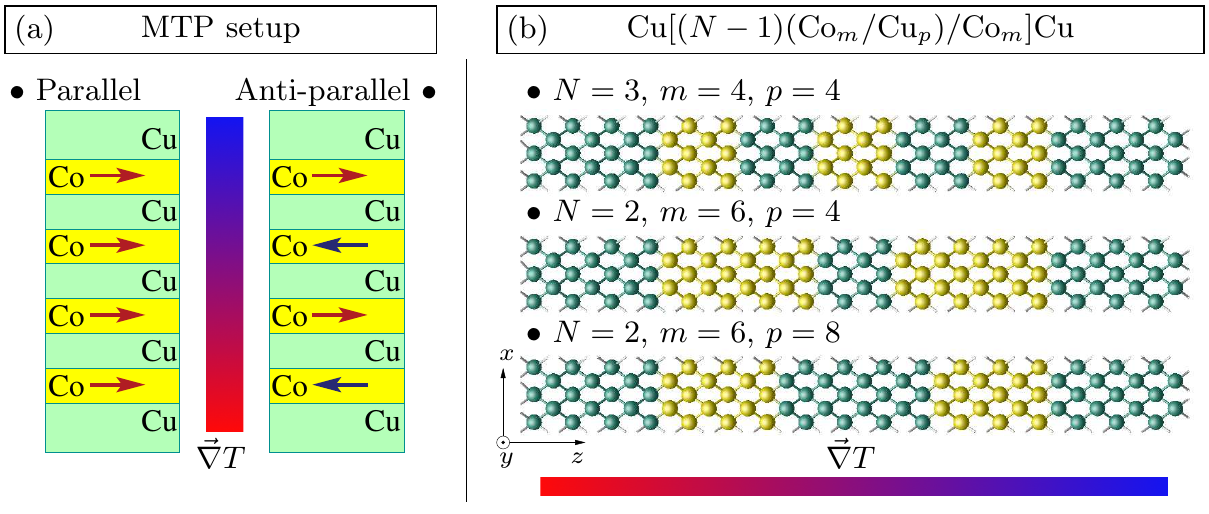}
  \caption{Schematic representations of (a) the generic setup of 
    a multilayered Co/Cu system providing an MTP signal; and (b)
    selected structural models of
    the Cu$[(N-1)$(Co$_m$/Cu$_q$)/Co$_m]$Cu systems investigated in this
    work, illustrating the meaning of the various geometrical 
    parameters $N$, $m$, and $q$. 
    $N$ is the number of Co repeats embedded in Cu, 
    while $m$ and $q$ represent the thickness in 
    atomic monolayers (MLs) of the Co
    (yellow) and Cu (dark green) layers. 
    The figure only shows the scattering region, with the half-infinite 
    Cu leads extending left and right along the $z$ direction.  
    The whole system is periodic in the $(x,y)$ plane. The thermopower 
    is calculated  along the temperature gradient which is taken to be 
    perpendicular to the interface.}
  \label{MTPandSLmodels}
\end{figure*}

In recent years, the ability to fabricate multilayer samples in the
form of nanopillars has opened up the possibility to detect their
internal state of magnetisation. The small diameter of the pillars, 
resulting in a small thermal conductance, in conjuction with a
strong heating by pulsed laser illumination of the pillar top 
allows one to build up sizeable temperature gradients 
\cite{GFR+04,GSGR+06,BVM+13,BNM+14,HKL+14}, causing a thermoelectric
voltage whose magnitude and sign reflects the internal 
magnetisation state. This method of detection,
depicted schematically in \Fref{MTPandSLmodels}(a),
may be even more sensitive than the CPP-GMR effect. 
Indeed, the magnetic response of a GMR device is usually quantified
through the GMR ratio, expressing the relative difference
of the resistances of the heterostructure in the two magnetic
alignments, parallel (P) and anti-parallel (AP). Equivalently,
one can use the conductance $g$ as defining quantity to
express the magneto-conductance (MC) ratio as:
\begin{equation}
  \label{MCrat}
  \mbox{MC} (\%) = \frac{g_{\rm P} - g_{\rm AP}}{g_{\rm P}} \,\times
  100 \enspace.
\end{equation}
The MTP ratio can be introduced quite analogously:
\begin{equation}
  \label{MTPrat}
  \mbox{MTP}(T) (\%) = \frac{S_{\rm P}(T) - S_{\rm AP}(T)}
                        {S_{\rm AP}(T)} \,\times  100 \enspace,
\end{equation}
using the temperature dependent
Seebeck coefficients for the two magnetic 
configurations $S_{\rm P}(T)$ and $S_{\rm AP}(T)$. 
Since $S_{\rm AP}(T)$ and $S_{\rm P}(T)$ may differ not only in
magnitude, but also in sign, one can imagine that the magnetic
contrast in a thermoelectric measurement, 
as expressed by the MTP ratio, may become 
larger than the MC ratio for a specific sample. 
Such an expectation could be confirmed
experimentally, for example by Gravier \etal~\cite{GSGR+06}.
In multilayered Co/Cu nanowires these authors found
an MTP ratio of $-30$~\%, larger than the $20$~\% measured 
GMR ratio. 
This behaviour is usually traced back to
the fact that the conductance 
(equivalently, the conductivity $\sigma$) 
is essentially a Fermi surface related property. 
The thermoelectric voltage (or the Seebeck coefficient),
on the other hand, is a measure of the energy dependence
of the relaxation rate near the Fermi energy $E_{\rm F}$
\cite{SYPS93}. 
As expressed by Mott's formula \cite{MJ58}, 
the Seebeck coefficient is proportional to the
logarithmic derivative of $\sigma(E)$:
\begin{equation}\label{MottEQ}
S = -\frac{\pi^2}{3e}k^2_{\rm B}T 
     \left.\frac{\rmd\ln\sigma(E)}{\rmd E}\right|_{E=E_{\rm F}}
    \enspace,
\end{equation}
where $e$ and $k_{\rm B}$ are, respectively, 
the elementary charge and the Boltzmann constant. On its basis,
one could derive a rather simple relation between the two 
quantities, the MTP and the MC ratios \cite{GFR+04}.

Phenomenological models, while being useful
in identifying general trends, do miss the important link between
the described quantity and the underlying electronic structure.
Precisely the opposite philosophy is adopted
in first-principles based investigations, as the ones presented
here: perform appropriate modifications of the electronic structure
and track the evolution of a given property with the ultimate
purpose of achieving specific design rules for a desired
target value.

For this purpose, we have considered one of the GMR prototypes,
the Co/Cu multilayered system. Many of its 
ground-state properties as well as the CIP- or CPP-GMR effects
have been already addressed on an {\em ab initio} level
\cite{KDTW94,NLZD94,KSK95,LNW+96,BZN+96,WBA+96,SUB+97,XZT+06,BZM00,BSW+02,SW05},
In contrast, first-principles calculations of
the magneto-thermoelectric properties of several
Co/Cu heterostructures, that require a significantly 
larger computational effort, 
gained only recently an increased attention \cite{PK13,KMWB14}.

The multilayered structure subject to our investigations can be
seen as a stacking of Co$_m$/Cu$_q$ bilayers of 
thickness $m$ and $q$ embedded in Cu(001). 
Accompanying the variations in the metallic layers thickness, 
the number of repeats $N$ was also treated as a 
variable, leading to the actual configuration
Cu[$(N-1)$(Cu$_m$/Cu$_q$)/Co$_m$]Cu, as schematically
shown in \Fref{MTPandSLmodels}(b). Note that, by
construction, $N$ was taken {\em finite}, that is,
no periodic boundary conditions along the
(001) growth direction were imposed.
The transport properties of these
systems are investigated
by performing first-principles 
calculations of the underlying electronic structure
by means of a spin-polarised relativistic Green's function
method \cite{PEP+04,PEP+05,EKM11}. The results obtained 
for the conductance and the Seebeck coefficient in a CPP geometry
|the temperature gradient taken perpendicular to the interface|
are analysed in view of the modifications in the electronic
structure induced by varying the 
morphology of the heterostructure, either through the number of 
repeats $N$ or of the thickness $m$ ($q$) of the
constituent Co (Cu) layers. 

The close lattice match of Co and Cu, as well as the advanced
fabrication technique of the nanopillars by electrodeposition, allows
the experimentalists to build stacks with a large number of Co
repeats. If the dominating scattering mechanism of the electrons is
scattering by the Co/Cu interfaces, it is to be expected that the
resistivity of a stack increases with the number of
repeats $N$, while the MC ratio is almost independent of $N$.  
For the Seebeck coefficient, which has the physical meaning 
of a voltage, its dependence on $N$ is not obvious. Our
calculations show that both $S(T)$ and the MTP ratio converge
rather rapidly with the number of Co repeats in the Co/Cu
stacks, reasonably converged values being attained already at $N=4$.

A modulation of the electronic density of states due to quantum 
confinement effects in ultra-thin layers may affect the
resistivity, but to an even higher degree the Seebeck coefficient 
of multilayered structures. The minority spin quantum well 
states formed in thin Co layers lie at the origin of 
an oscillatory behaviour observed for 
many physical properties of these systems, ranging from
the interlayer exchange coupling \cite{KDTW94,NLZD94}
or the magnetic anisotropy energy \cite{SUB+97}
to the recently investigated Seebeck magnetic anisotropy \cite{PK13}.
Our calculations show that these quantum well states hybridise
with a high-mobility band crossing the Fermi energy.
As a result, we find strong variations in amplitude and sign of 
both the Seebeck coefficient and the MTP occurring 
in multilayers consisting of Co stacks of up to seven monolayers.
For the design of Cu/Co stacks for MTP read-out, this would mean that 
depositing a small number of Co layers with precise control of the
layer thickness is more useful than
increasing the number of Co repeats in the stack. By comparison,
the Co and Cu thickness dependence
of conductance and MC ratio was found to be much weaker.

The paper is organised as follows: We start by providing the relevant
computational details, including the geometry of the systems,
in section~\ref{SecCompute}. Section~\ref{CoSingleLayer} deals with
the particular case of a single Co layer embedded in Cu(001), in which
we mostly focus on aspects of the electronic structure, with a
particular attention paid to the quantum well states appearing in
the Co layer. Last two sections are devoted to the transport
properties of the 
Cu[$(N-1)$(Cu$_m$/Cu$_q$)/Co$_m$]Cu multilayers, with detailed
discussions on their $N$, $m$, and $q$ dependence. 

\section{Geometry of the system and 
theoretical background}\label{SecCompute}

The calculations for the Cu$[(N-1)$(Co$_m$/Cu$_q$)/Co$_m]$Cu 
multilayer systems were performed using a spin-polarised
relativistic (SPR) \cite{EKM11,SPR-TB-KKR} version of the
screened Korringa-Kohn-Rostoker Green's function (KKR-GF)
method \cite{SUWK94,ZDU+95,WZD97}. We apply the same
procedure as described in our previous investigations on 
the Cu/Co$_m$/Cu trilayers \cite{PK13}, that essentially consists
of three steps: (i) setting up the geometry of the system;
(ii) the self-consistent determination of the ground state 
potentials; and (iii) using these as input for
the transport calculations which are based on
the Landauer-B\"uttiker formula as implemented in 
the KKR-GF method \cite{BS89,MPD04} within a relativistic
representation \cite{PEP+04,PEP+05}.
Our approach, discussed to some
extent in this section, has the 
one-electron retarded Green's function 
$G^+(\vec r,\vec r{\,'}; \varepsilon)$ at energy 
$\varepsilon = E + \rmi\delta$ as central quantity.

\subsection{Modelling the multilayered systems}

We model the systems under investigation 
by taking two half-infinite Cu leads with an interaction 
region inserted in-between, all sharing the same in-plane two 
dimensional (2D) periodic lattice. 
Since the natural lattice misfit between elemental 
Co and Cu is rather small (less than $2$~\%), 
we neglect the lattice relaxation at the interfaces 
and take all atomic positions as being fixed to the ideal
(001)-stacked fcc lattice with the lattice constant equal to the
experimental fcc-Cu value of $3.61$~\AA. 
The interaction region contains the 
$[(N-1)$(Co$_m$/Cu$_q$)/Co$_m]$ multilayered structure
and up to $10$  atomic monolayers (MLs)
of Cu on its both sides. These additional Cu MLs
are meant to ensure a smooth transition towards the Cu leads.

Schematic representations of selected
setups for the interaction region
are provided in \Fref{MTPandSLmodels}(b) for
varying number of Co repeats (here, $N=2$ and $N=3$) and 
individual Co ($m=4$ and $m=6$~ML) and Cu 
($q=4$ and $q=8$~ML) layer thickness. For our investigations we
had considered, changing just one variable at a time, 
$N=1,\ldots,6$ and combinations of $m$ and $q$
ranging between $4$ and $8$~MLs thickness.
We have determined the longitudinal thermopower occurring under a
temperature gradient taken parallel to the growth direction $z$. 
Note that, along this direction,
no periodic boundary conditions are imposed. 
We furthermore emphasise that 
different numbers of repeats
effectively mean differently sized finite objects along $z$;
an increase in $N$ (at a given $m$ and $q$)
is equivalent to an increase in the
thickness of the interaction region.

\subsection{Electronic structure calculations}

For each of the configurations, the potentials are determined
self-consistently using the screened KKR-GF method \cite{EKM11},
considering spherical potentials in the atomic
sphere approximation (ASA) 
within the local spin-density approximation in the
Vosko, Wilk and Nussair parametrisation \cite{VWN80}. 
An angular momentum cut-off of $l_{\rm max}=3$ was taken for the Green's 
function expansion.

In a preliminary step, a separate self-consistent calculation
is performed in order to determine the potential of the
two (identical) half-spaces left and right of the interaction
region. A second self-consistent
procedure is applied to the interaction region itself, in which
all its potentials are iterated, whereby the outer-most Cu
potentials asymptotically match the ones in the leads.
This matching is accomplished by means of the decimation
technique \cite{TDK+97}, in which the leads potentials determined
in the first step provide the appropriate boundary conditions
of the heterostructure. Different magnetic couplings between adjacent
Co layers, parallel and anti-parallel, 
were separately considered at each $(N,m,q)$ combination. The
collinearity of the spin magnetic moments was the
only constraint imposed {\em a priori}.

As a consequence of the 2D-periodicity of the layered system, the Green's
function can be Fourier transformed in a 2D representation with the
Bloch vector $\vec k_\parallel$ as constant of motion and retaining an
index $i$ for the position along the growth direction $z$.
Within the KKR-GF scheme, the Green's function is
expressed in terms of the matrix 
$\mat{G}^{ij}(\vec k_\parallel,\varepsilon)$.
This matrix describes the propagation of the electron wave
between the atomic sites $i$ and $j$ at positions 
$\vec R_i$, $\vec R_j$ and is labelled,
in our adopted representation, by the relativistic quantum numbers
$\Lambda=(\kappa,\mu)$, i.e. 
$(\mat{A})_{\Lambda\Lambda'}=A_{\Lambda\Lambda'}$ \cite{EKM11}.
Let us note here that, since the spin is not a constant of
motion we shall use the designation majority/minority spin 
rather than up/down ($\uparrow/\downarrow$). 

\subsection{Transmission probability,
conductance  and the Seebeck coefficient}

Combining the structural Green's function matrix 
calculated for a given 2D-periodic system with the 
matrices $\mat{M}^i$, $\mat{M}^j$ of the
$z$-component of the relativistic current operator at sites
$i$ and $j$ enables the calculation of the electronic
transmission probability between two atomic planes $I$ and $J$ 
according to the expression \cite{PEP+05}:
\begin{equation}
  \label{TkparE}
  {\cal T}(\vec k_\parallel,E) =
  \sum_{i\in I,j\in J} {\rm Tr} \left[
     \mat{M}^{i\dagger}\mat{G}^{ij}(\vec k_\parallel,\varepsilon)
     \mat{M}^{j}\mat{G}^{ij\dagger}(\vec k_\parallel,\varepsilon)\right]\enspace,
\end{equation}
where each 2D vector $\vec k_\parallel$ can be seen as a conduction
channel \cite{MPD04}. By integrating over the 2D Brillouin zone (2D-BZ)
the total transmission probability ${\cal T} (E)$ at energy $E$
is then \cite{MPD04}:
\begin{equation}  \label{TkINT}
 {\cal T} (E) = \frac{1}{A_{\rm 2D-BZ}}\int_{\rm 2D-BZ}\!\! \rmd^2\vec
 k_\parallel {\cal T}(\vec k_\parallel,E)\enspace.
\end{equation}

In the case of a weak spin-orbit coupling, as it is the case for the
light 3d transition metals, Popescu \etal \cite{PEP+04,PEP+05}
could show that the transmission
through the ''fully relativistic resistor'' expressed by
\Eref{TkparE} can be approximated by 
$\widetilde{\cal{T}}k_\parallel,E)$:
\begin{eqnarray}\nonumber
{\cal T}(\vec k_\parallel,E) \simeq  \widetilde{\cal{T}}(\vec
  k_\parallel,E) & = & 
  \widetilde{\cal{T}}_{\uparrow\uparrow}(\vec k_\parallel,E) +
  \widetilde{\cal{T}}_{\downarrow\downarrow}(\vec k_\parallel,E) + \\
  & &
  + \widetilde{\cal{T}}_{\uparrow\downarrow}(\vec k_\parallel,E) +
    \widetilde{\cal{T}}_{\downarrow\uparrow}(\vec k_\parallel,E)
  \label{TSpinDec}
\end{eqnarray}
which provides a spin decomposition 
essentially equivalent to the Mott two-current model. 
\Eref{TSpinDec} can be regarded as its generalisation
to the relativistic case. In addition to the spin-conserving
($\uparrow\uparrow$ and $\downarrow\downarrow$)
channels, it also includes 
spin-mixed ones ($\uparrow\downarrow + \downarrow\uparrow$),
induced by the spin-orbit coupling.
We will use this approximate spin decomposition only for
a qualitative discussion in section~\ref{CoSingleESandT}.

Following Sivan and Imry \cite{SI86}, 
the Seebeck coefficient $S(T)$
can be obtained from ${\cal T} (E)$ through the expression
\begin{equation}
 \label{SEEBECK}
  S(T)  = -\frac{1}{eT} \,
        \frac{{\displaystyle \int {\rm d}E \,
             \partial_E f_0 \, {\cal T}(E)}
             \,(E-E_{\rm F})}
             {{\displaystyle\int {\rm d}E \,
         \partial_E f_0 \, {\cal T}(E)}}\enspace,
\end{equation}
where $f_0\equiv f_0(E,T,\mu)$ is the Fermi-Dirac distribution function at
energy $E$, temperature $T$, and chemical potential $\mu$,
while $\partial_E f_0=\partial f_0/\partial E$ represents
its energy derivative. The denominator in the last equation is
related to the temperature dependent conductance $g(T)$ by:
\begin{equation}
 \label{CONDUCT}
  g(T)  = -\frac{e^2}{h} \,
        \int {\rm d}E \, \partial_E f_0 \,
        {\cal T}(E)\enspace.
\end{equation}

The various parameters involved in the actual evaluation of these
quantities were chosen in the following way: For the 2D-BZ integral 
required for the transmission probability, \Eref{TkINT}, a regular 
$1000\times 1000$ $\vec k_\parallel$-grid was found necessary
to achieve convergency of ${\cal T}(E)$
over a broad range of energy arguments.
For the integrals in \Eref{SEEBECK} and \eref{CONDUCT}, 
on the other hand, ${\cal T}(E)$ was explicitly calculated on
a $1$~mRy-spaced regular mesh, then interpolated on a denser
mesh of $0.1$~mRy. In these equations, the limits of the 
energy interval below and above $E_{\rm F}$
were set in such a way that 
$\partial_E f_0(E_{\rm min/max})<10^{-8}$, a limit 
found to be more than sufficient in providing well-converged
results.

\subsection{Explicit temperature dependent effects}

The formalism employed here rigorously describes
elastic scattering at the interfaces and treats the simultaneous
occurrence of spin polarisation and relativistic effects,
such as spin orbit coupling, on equal footing. 
Temperature enters in this approach through the Fermi-Dirac 
distribution function, but temperature-dependent scattering, 
e.g. by  atomic vibrations or spin fluctuations are neglected.

Inclusion of atomic displacements at finite temperature in transport properties 
calculations within {\em ab initio} methods has been recently 
accomplished by treating them as static disorder via the coherent 
potential approximation (CPA) \cite{KCME13}. 
Alternatively, one could use large 2D supercells and apply a frozen 
phonon approach averaging over explicit different atomic displacements.
We note, however, that $S(T)$, being the quotient of two integrals
involving the transmission probability ${\cal T(E)}$,
any additional temperature dependence due to inelastic scattering,
appearing both in the numerator and the denominator, tends to cancel
out as long as phonon drag effects can be disregarded.

Accounting for 
electron scattering by spin fluctuations in various 
ferromagnetic metals and alloys has been convincingly 
demonstrated to improve the agreement between calculated 
and experimentally determined temperature dependent 
resistivity \cite{WSvsB09,KDT+12}. More recently, 
Kov\'a\v{c}ik \etal\ \cite{KMWB14} investigated
the effect of {\em static} spin disorder on the magneto-thermoelectric 
phenomena of several nano-structured Co/Cu systems. 
These authors could show that, while the spin-dependent 
electron scattering does indeed influence the spin-caloric transport 
coefficients at elevated temperatures, the way in which it
manifests itself is strongly case dependent. In particular,
no general trends could be identified, 
neither do quantitative nor qualitative predictions 
appear to be possible without an explicit calculation \cite{KMWB14}. 
To what extent {\em dynamic} spin fluctuations 
influence the Seebeck coefficient is hardly explored. 
It may be noted  that Piraux \etal~\cite{PFS+92}  invoked 
inelastic spin-dependent electron-magnon scattering to explain the
increase in the MTP at high temperatures observed experimentally in
Co/Cu and Fe/Cu thin multilayers. However, as we will show below, an equally
large MTP can be obtained accounting solely for the electronic
structure contributions to the thermopower. 
For systems with a gapped band structure, such as magnetic half-metals
or tunnel junctions, 
modifications of the electronic properties due to dynamic spin
fluctuations at finite 
temperatures have been addressed via phenomenological 
models \cite{MN14,IK06} or via the dynamical mean field 
theory (DMFT) \cite{CSA+08}. 
For systems as large as those considered here, such an advanced
many-particle treatment is 
computationally not feasible at present.

The importance of these temperature dependent effects notwithstanding,
our primary focus here is to identify the specific effects
on the magneto-thermopower which are 
intimately connected with the electronic structure and 
are solely induced by quantum confinement. 
As such, properly accounting for the effects
discussed above is well beyond the purpose of the current 
investigations, although this is clearly needed in future 
studies for an improved quantitative agreement with experiment.

\section{A single Co layer embedded in Cu(001)}
\label{CoSingleLayer}

We begin our discussion by presenting results obtained for a single Co
layer embedded in Cu(001), a geometry setup corresponding to $N=1$
in the general notation Cu$[(N-1)($Co$_m$/Cu$_q)$Co$_m]$Cu introduced
above. We shall first analyse briefly the electronic structure in the
proximity of the Cu/Co interface and illustrate how its peculiarities 
are reflected in the transmission probability for a single Co layer 
of varying thickness $m$. The ${\cal T}(E)$ transmission profile for
$m=4$ will be shown to exhibit a peak in the minority spin channel
immediately below the Fermi energy. This peak 
is intimately connected with a complex formed by 
a quantum well state (QWS) and a high mobility p-band present
in the interface layers, with which the QWS hybridises. These findings
for the single layer system will be important in understanding the
transport properties of the multilayered
Cu$[(N-1)($Co$_m$/Cu$_q)$/Co$_m]$Cu systems.

\subsection{Density of states and 
              transmission probability}\label{CoSingleESandT}

\begin{figure*}[t]
  \centering
  \includegraphics[width=0.85\textwidth]{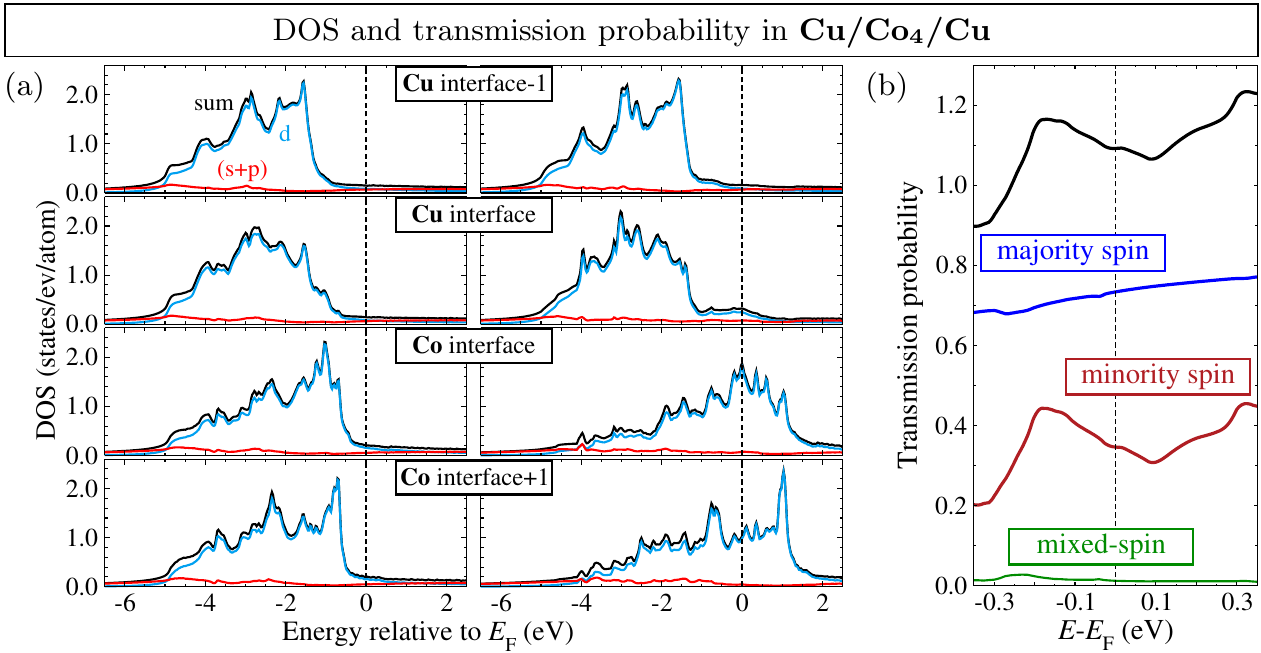}
  \caption{
    (a) Spin- and angular momentum resolved local DOS projected on the 
    Cu and Co atoms in the vicinity of the Cu/Co interface, with the
    majority (minority) spin channel in the left (right) panels.
    The (s+p)-DOS are shown as sum (red lines) while the
    d-DOS (light blue) is seen to have the dominant contribution
    to the total DOS (black).
    (b) Electronic transmission probability in 
    the Cu/Co$_4$/Cu trilayer system. The total transmission (black)
    is decomposed in its spin-conserving (blue and dark red) 
    and spin-mixing (green) components. Note the pronounced peak in
    the minority spin transmission at about $-0.2$~eV. The Fermi
    energy, set by the Cu leads, is taken as reference in this figure.}
  \label{DOSandTRANSFig}
\end{figure*}

Electronic structure calculations performed on the Co/Cu systems
\cite{KDTW94,BZN+96,BSW+02,SW05} revealed that
the majority spin d-band is completely filled and
the energy range at and near the Fermi level
is dominated by the 3d minority spin states stemming from Co.
These features are accordingly reproduced by our calculations
and reflected in the spin-resolved density of states (DOS) 
for the Cu/Co$_4$/Cu trilayer system shown in
\Fref{DOSandTRANSFig}(a). Here the DOS is 
further projected on the Co and Cu atoms in the vicinity of
the Cu/Co interface as well as on their angular momentum
(s+p)- and d-components.

\Fref{DOSandTRANSFig}(a) evidences that the d-DOS 
(light blue curve) of both Co and Cu has an overwhelming contribution 
to the total DOS of the Co/Cu heterostructure in both spin 
channels. We further note the large Co-related contribution in the
minority spin channel in the proximity of the Fermi energy, 
contrasting the extremely reduced DOS on the Cu sites. Only the
first Cu layer near the interface exhibits a slight spin polarisation,
induced by its neighbouring Co atoms. 

The dominant d-character of the minority spin states close
to the Fermi energy is accordingly reflected in the 
transport properties. Indeed, the d-states are characterised by a
stronger localisation and a reduced mobility as compared
to the s- and p-states. We show in \Fref{DOSandTRANSFig}(b) 
the transmission probability ${\cal T}(E)$
for the same Cu/Co$_4$/Cu trilayer system, containing a 4~ML thick Co
layer. The black curve in this figure represents the total 
transmission, calculated for various energy arguments $E$ using
\Eref{TkINT}. Applying the spin decomposition leading to the
approximate form \eref{TSpinDec}, allows us to identify the
spin-conserving and spin-mixing transmission channels in
${\cal T}(E)$. As a result of the rather small spin-orbit
coupling, the spin-mixing transmission (green lines) is negligibly
small. In spite of a small DOS near $E_{\rm F}$, 
the largest transmission component is the spin-conserving 
majority one. The minority spin transmission is only about one 
third of the total. Nevertheless,
while $\widetilde{{\cal T}}_{\uparrow\uparrow}(E)$ 
is nearly featureless and shows a weak variation with $E$, it is
the $\widetilde{{\cal T}}_{\downarrow\downarrow}(E)$ 
component which modulates the full transmission profile 
${\cal T}(E)$. This different
qualitative behaviour arises from the different character of the
states involved in the transmission through the various channels:
nearly exclusively s- and p-states for the majority spin, 
hybrid s-p- {\em and} d-states for the minority spin. 
The first important conclusion of our investigations can thus
be formulated as follows: 
Although comparatively small in magnitude, the minority spin 
transmission is expected to be much more sensitive to the 
morphology of the system. Changes in the transport properties
caused by geometry modifications can be traced back nearly exclusively
to modifications in the minority spin electronic structure.

While the general characteristics of the transmission profile 
discussed above were found to be valid for all the systems
investigated, the pronounced peak in the minority spin transmission
[dark red line in \Fref{DOSandTRANSFig}(b)],
$0.2$~eV below the Fermi energy is a peculiarity of
the chosen Co thickness, $m=4$~ML. It represents, in fact, a 
signature of a QWS arising in the minority spin band of Co. 
The detailed discussion of these QWSs 
makes the subject of the next section.

\subsection{Quantum well states}

The appearance of QWSs in the Co slab has been found responsible
for the non-monotonous behaviour of the magnetic 
anisotropy energy (MAE), evidenced
both experimentally \cite{WBA+96} and theoretically \cite{SUB+97}
in the Cu/Co$_m$/Cu trilayers, as well as for the 
oscillations in the interlayer exchange coupling 
occurring in Co/Cu for thin Co layers \cite{LNW+96}.
In a recent study of the authors \cite{PK13}, the QWSs were also 
shown to play an important
role in inducing an anisotropic MTP
in the same systems. In particular, we identified a hybrid 
complex formed by the QWSs and a p-type band specific to the Co/Cu
interface. These hybrid states provide extremely efficient channels
for the minority spin electrons, as evidenced by the aforementioned
peak, $0.2$~eV below $E_{\rm F}$, in Cu/Co$_4$/Cu. In the next
sections we shall link our findings for the Seebeck coefficient 
and the MTP precisely to these peculiarities of the electronic
structure.

The minority spin channel QWSs appearing in the 
Co slab have been investigated by calculating the angular momentum 
and atom projected Bloch spectral function (BSF)
$A_i(\vec k_\parallel,E)$ \cite{EKM11}:
\begin{eqnarray}
  A_i(\vec k_\parallel,E) & = & -\frac{1}{\pi N}\,{\rm Im}\,{\rm Tr}
         \sum_{n,n'}^N \rme^{\rmi\vec k_\parallel(\vec
         \chi_n-\vec\chi_{n'})} \nonumber\\
      & & \times   \int\rmd^2r_\parallel 
         G(\vec r_\parallel+\vec R_i+\vec\chi_n,
           \vec r_\parallel+\vec R_i+\vec\chi_{n'};E)\enspace,
        \label{BSFkpara}
\end{eqnarray}
where $\vec\chi_n$, $\vec\chi_{n'}$ are translation vectors of the
2D periodic lattice, $\vec r_\parallel=(r_x,r_y,0)$,
and $\vec R_i=(0,0,z_i)$ represents the $z$-coordinate of the
$i$th atom. The BSF is a quantity that can be regarded as a 
$\vec k_\parallel$-resolved DOS. 

\begin{figure*}[t]
  \centering
  \includegraphics[width=0.96\textwidth]{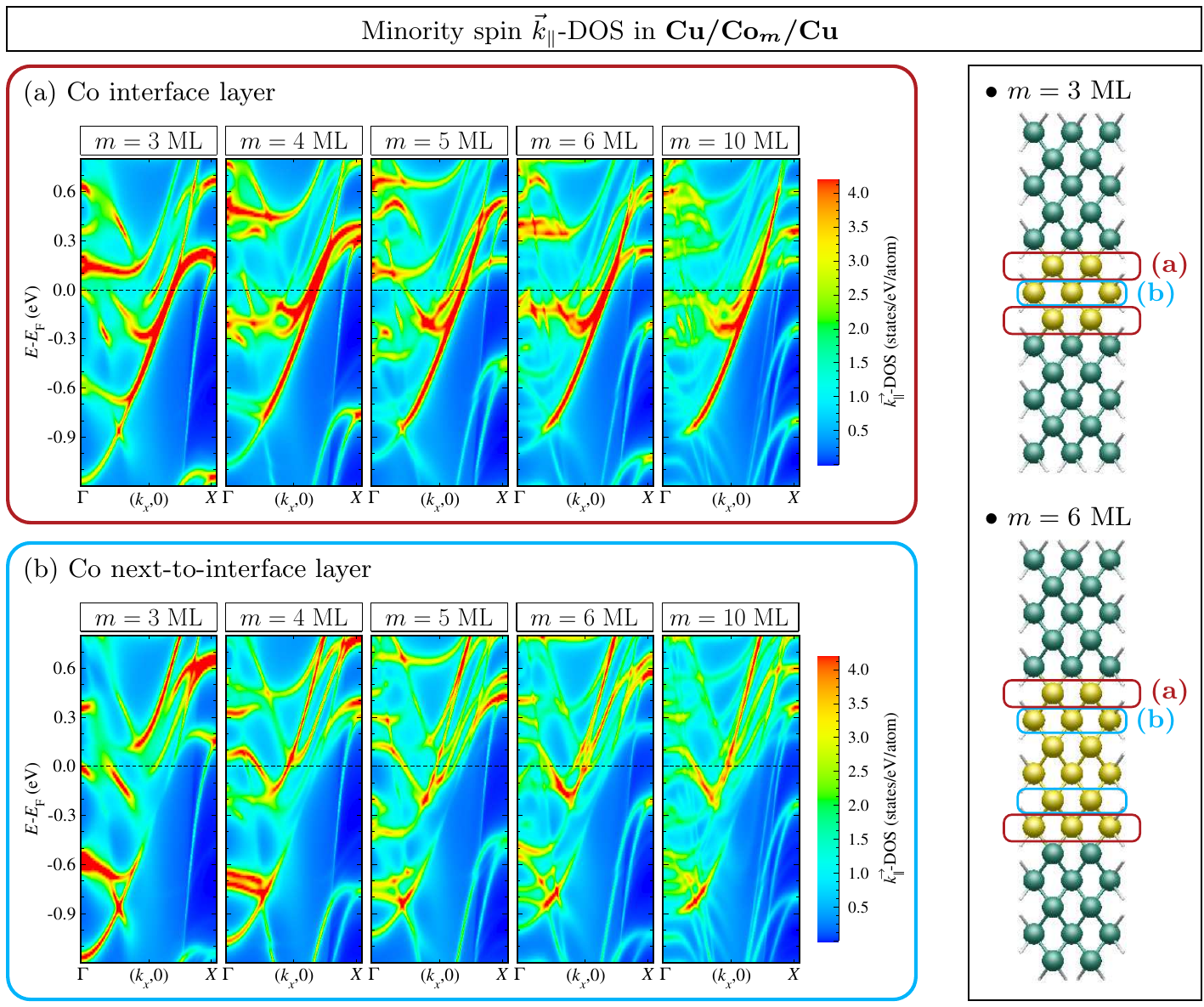}
  \caption{
    Minority spin $\vec k_\parallel$-resolved DOS projected on
    the (a) interface and (b) next-to-interface Co atom
    in the Cu/Co$_m$/Cu trilayer system
    for various values of $m$, as shown schematically on the
    right side of the figure. 
    For the interface layer (a)
    typical signatures of QWSs appearing as flat bands
    above (odd $m$) and below (even $m$)
    $E_{\rm F}$ for thin Co slabs as well as 
    a p-type band crossing the
    Fermi energy $E_{\rm F}$ (taken here as reference value) 
    can be recognised. The latter has only a weak correspondent
    in the next-to-interface layer (b). Note that the same scale has
    been used in both panels.}
  \label{QWSFig}
\end{figure*}

Figure~\ref{QWSFig} depicts the minority spin
component of $A_{\rm Co}(\vec k_\parallel,E)$ projected on the first
[panel (a)] and second [panel (b)] 
Co atomic layers near the Cu/Co interface,
with the respective location of each layer
schematically drawn at the side of the figure. 
The different frames in each panel follow the
variation of the Co layer thickness $m$~MLs in the
Cu/Co$_m$/Cu trilayer system.
Such $E$ versus $\vec k_\parallel$-type plots allow 
us to identify the projected band structure in the 2D-BZ, a picture 
familiar from angle-resolved photo-emission experiments.

The formation and appearance of a certain QWS will depend on the
Co thickness $m$, alternating between odd and even number of MLs. 
For a given parity, on the other hand, the $m$-dependence is reflected in
a variation in the energy position of the QWS. 
Typical signatures of QWS can be observed as flat bands
in the BSF of the interface Co layer [panel (a)]
near the 2D-BZ centre: (i) around $0.15$~eV 
for $m=3$ and $m=5$ and (ii) around $-0.2$~eV and $0.4$~eV
for $m=4$ and $m=6$, energy values relative to the Fermi level.
With increased thickness of the Co slab, the QWSs morph into a
continuum, as seen in the right-most frame of \Fref{QWSFig}(a)
for $m=10$~MLs.

The second important aspect revealed by the top panel of 
\Fref{QWSFig} is the existence of a high-mobility p-band
for the minority spin carriers, crossing the Fermi energy.
This band, evidenced by the red-coloured, 
$S$-shaped feature of the BSF in \Fref{QWSFig}(a), 
stems from the Cu and Co atoms adjacent 
to the interface and exhibits no thickness dependence.
In other words, it is an ubiquitous characteristic of the 
Co/Cu interface. 

For the case of an {\em even} number of MLs $m$, 
the QWS forming below the Fermi energy will couple to this p-band, 
leading to the formation of a p-d hybrid complex. This is precisely 
the origin of the strong transmission evidenced in the minority 
spin channel at $-0.2$~eV in \Fref{DOSandTRANSFig}(b) and, 
as shown below, for the positive value of the Seebeck coefficient 
associated with $m=4$~ML Co thick systems. Note that a similar 
hybrid band complex also appears above the Fermi energy, at $0.45$~eV.
This is however too far to contribute to the integrand of 
\Eref{SEEBECK}.

It is easy to see how the morphology of the system
may have significant influences on its electronic structure, and, as a
result, on its transport properties. By comparing
the spectral functions of the two different Co atoms in
\Fref{QWSFig}(a) and (b) one can see,
for the next-to-interface layer, 
a significant reduction in the amplitude of 
the p-band-related BSF below the Fermi energy. 
Thus, for the transmission channels opened 
by the QWS-p-band complex it will mean that their weight and 
importance in the total transmission will diminish with increasing 
Co thickness. This effect is further amplified by the smearing out
of the QWSs. 
When $m$ is {\em odd}, on the other hand,
the QWS appearing above the Fermi energy couples very weakly, 
if at all, with the highly mobile p-band. As a consequence, no 
corresponding high energy transmission peak is expected for odd $m$.
This thickness-triggered filtering of the transmission channels
involved in the conduction was found responsible for the
non-monotonous behaviour of the Seebeck coefficient in 
Cu/Co$_m$/Cu trilayers \cite{PK13}. 

Additional variables come into play in the case of
multi-layered systems: The size of the Cu spacer placed between the Co
layers will modify the way in which the different states
will or will not couple across the interfaces, leading to an
enhancement or suppression of various transmission channels. Likewise,
a varying number of Co and Cu repeats may further complicate the
picture. Furthermore, when different magnetic alignments
between adjacent Co layers are considered, 
one has to bear in mind that the
minority/majority spin channels get swapped.
We discuss these aspects in the next sections, essentially
showing that a broad range of values may arise for both the Seebeck
coefficient and the MTP, depending on the different 
morphological parameters. 

\section{Varying number of Co repeats}

In the previous section the discussion focused on
electronic structure characteristics related to a single 
Co layer of varying thickness embedded in Cu(001). We have
analysed how these may influence directly the transmission 
probability and, through it,
the various transport properties. The first question to ask is
how much and to what extent the knowledge gained so far is
transferable to the multilayered
Cu$[(N-1)$(Co$_m$/Cu$_q$)/Co$_m]$Cu systems. 
In this section we discuss results obtained by modifying 
the number of Co repeats $N$ while keeping the other parameters,
$m$ and $q$, fixed. For convenience and easier comparison with
the results already presented, we restrict the discussion,
without losing any generality, to the case $m=q=4$~MLs. 
The most important
conclusions drawn at the end of this section are:
(i) the electronic structure features present in
the single-Co layer system transfer to the Co-stacked
systems; (ii) increasing the number of Co
repeats reduces the transmission through the
heterostructure without, however, significantly
modifying its $E$-dependent profile; and (iii) 
at high temperature, the
Seebeck coefficients, both for parallel and anti-parallel
alignments, as well as the derived MTP are reaching 
converged values in $N$ rather fast.

\subsection{Electronic structure and 
               transmission probability}

\begin{figure*}[t]
  \centering
  \includegraphics[width=0.7887\textwidth]{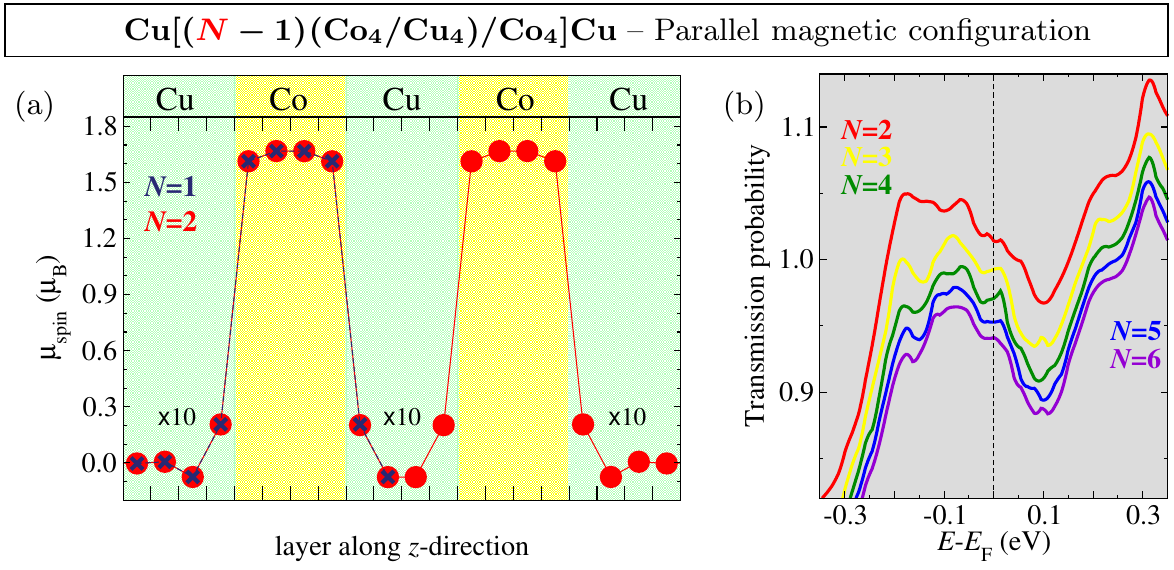}
  \caption{(a) Spin magnetisation profile of the 
    Cu/Co$_4$Cu$_4$Co$_4$/Cu system ($N=2$, red bullets) 
    with a parallel (P) magnetic configuration as compared
    to that of the Cu/Co$_4$/Cu trilayer ($N=1$, dark blue crosses).
    The figure emphasises on the quasi independence of the 
    spin magnetic moments in the Co layers on the number of repeats
    $N$. Note that the values corresponding to the Cu atoms are
    multiplied by a factor of $10$.
    (b) $N$-dependence of the electronic transmission probability
    ${\cal T}(E)$ 
    in Cu$[(N-1)$(Co$_4$/Cu$_4$)/Co$_4]$Cu with a
    P-alignment of the Co magnetic moments. The transmission curves
    are shown around the Fermi energy $E_{\rm F}$, taken here
    as reference in the same range as in \protect\Fref{DOSandTRANSFig}(b).}
  \label{NCoCuMUETRANS}
\end{figure*}

The electronic structure calculations performed for the
Cu$[(N-1)$(Co$_m$/Cu$_q$)/Co$_m]$Cu systems with $N=1,\ldots,6$
revealed an interesting feature: the calculated ground state
properties such as spin and orbital magnetic moments, DOS or BSF
curves projected on the individual components exhibit a rather weak
dependence on $N$. As an illustrative example we show in 
\Fref{NCoCuMUETRANS}(a) the spin magnetisation profiles for
$N=1$ (red bullets)
and $N=2$ (dark blue crosses), that is, one and two Co slabs 
embedded in Cu, each of a thickness $m=4$~MLs.
For $N=2$ the two Co layers are separated  by a thin
Cu spacer ($q=4$~MLs). One can see that there are hardly any 
differences noticeable in the individual spin magnetic moments
on the Co atoms in the two cases, 
Cu/Co$_4$/Cu and Cu/Co$_4$Cu$_4$Co$_4$/Cu. Specifically,
for the Co atoms nearest to Cu, the spin magnetic
moments obtained were (in Bohr magnetons $\mu_{\rm B}$):
$1.6137$ ($N=1$), $1.6134$ ($N=2$, outer Co),
and $1.6130$ ($N=2$, inner Co). 
In spite of the very small thickness of the spacer, 
the individual Co layers obviously display
the same properties, regardless of $N$. 
Analogous results were obtained for the
other number of repeats; furthermore, also the Cu spacer 
layers of equal thickness exhibit similar characteristics.

In other words, these findings imply that all the considerations made
in the previous section regarding the electronic structure of a single
Co slab transfer to the multi-layered systems,
qualitatively completely and quantitatively 
to a large extent. This also holds, in particular, for the
QWSs formation and the Co/Cu-interface specific p-band in the minority
spin channel.

Not surprisingly, a similar one-to-one transferability holds
only partly in the transmission probability profiles.
These are shown in \Fref{NCoCuMUETRANS}(b) for $m=q=4$~MLs, 
$N=1,\ldots,6$, and with all the magnetic moments oriented parallel 
one to another. Note that throughout the next figures we will use the 
same colour-coding convention,
borrowed from the solar spectrum: as the variable under investigation
increases, the colours used in the graphical representation 
change from red to purple. The general trend that can be recognised 
from \Fref{NCoCuMUETRANS}(b) is an overall down-scaling of the
${\cal T}(E)$ profiles with increasing $N$. This is a direct
consequence of successively adding interfaces, that is, electron 
scattering sources, to the transmission process
between the left and right leads. It is nevertheless obvious that the
reduction in transmission is not a uniform function of energy
argument. For this reason, the peak at $-0.2$~eV discussed above 
for $N=1$, although still present for all values of $N$, 
appears of varying shape and width. We also point out to the 
slope of ${\cal T}(E)$ near the Fermi energy, which is changing 
sign with $N$. While being the result of subtle variations 
in the way different states couple, these changes influence the 
thermoelectric properties of the multilayered system only in 
the limit of low temperatures.

\subsection{Spin-dependent thermally induced 
               transport properties}\label{SebCondVarN}

\begin{figure*}[t]
  \centering
  \includegraphics[width=0.8267\textwidth]{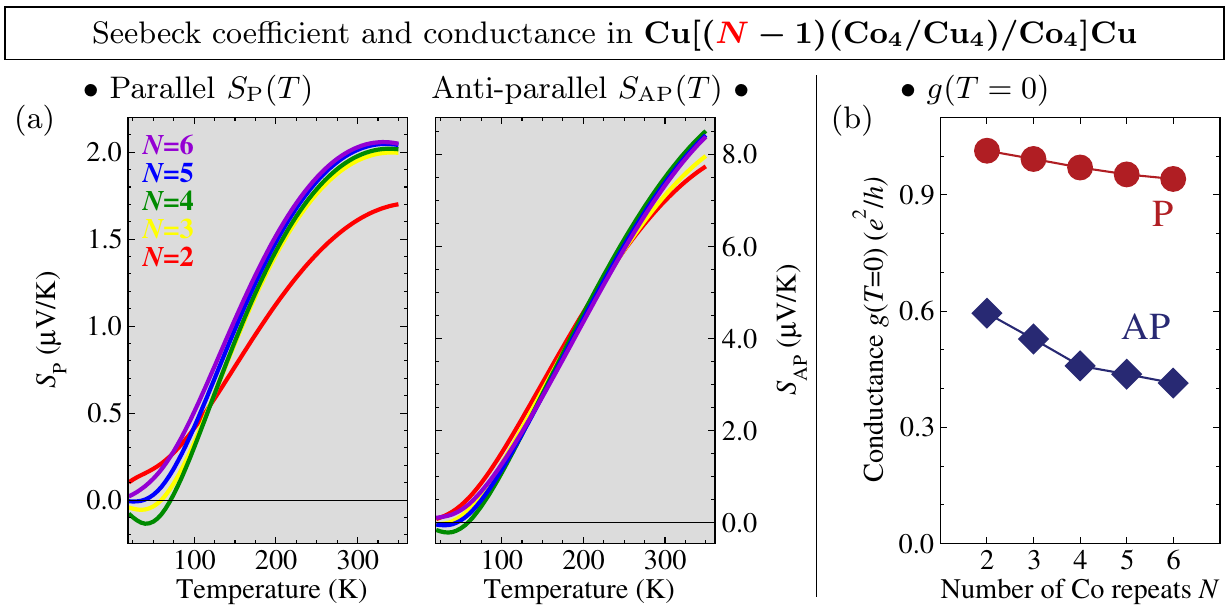}
  \caption{Dependence of (a) the
    Seebeck coefficient $S(T)$ and (b) the zero temperature 
    conductance $g(T=0)$ on the number of Co repeats $N$ in the 
    Cu$[(N-1)$(Co$_4$/Cu$_4$)/Co$_4]$Cu
    multilayered systems:
    (a) $S_{\rm P}(T)$ for parallel (P, left panel)
        and $S_{\rm AP}(T)$ for anti-parallel (AP, right panel)
       alignment of the magnetisation in the Co layers.
    (b) conductance for P (bullets) and AP (diamonds) alignments.
    Note the different scale used for $S_{\rm P}(T)$ and $S_{\rm AP}(T)$.}
  \label{NCoCuSeb}
\end{figure*}

\Fref{NCoCuSeb}(a) displays the dependence of the
Seebeck coefficient on the number of Co repeats $N$
in the multilayered Cu$[(N-1)$(Co$_4$/Cu$_4$)/Co$_4]$Cu systems.
The two panels correspond to results obtained 
for two magnetic configurations, 
a parallel (P) and an anti-parallel (AP) alignment.
The latter is understood as the succession of Co slabs, each of
$m$~MLs thickness (here $m=4$), in which 
the magnetic moments in one slab are opposed to 
those of its neighbouring Co slabs. 
As an example, for $N=5$ one would have for the Co
slabs the arrangement
($\uparrow\uparrow\uparrow\uparrow\uparrow$) for the P-alignment and
($\uparrow\downarrow\uparrow\downarrow\uparrow$) for AP.

As mentioned in the introduction, measurements of the Seebeck 
coefficient are usually interpreted on the basis of
Mott's formula, \Eref{MottEQ},
which provides a direct link between  $S(T)$ and the
conductivity $\sigma(E)$ of a sample. Its range of validity has been
discussed to some extent in the literature, e.g. by
Jonson and Mahan \cite{JM80}, who showed that it gives the
correct $T\to 0$ behaviour for independent electrons interacting 
with static impurities and for adiabatic phonons. 
Beyond Mott's formula, the way in which a transmission 
probability profile ${\cal T}(E)$ influences the sign and size of
the Seebeck coefficient at finite $T$ can 
be understood on the basis of \Eref{SEEBECK}.
In this equation, a temperature increase effectively 
extends the integration range, by increasing 
the non-zero width of ${\cal T}(E)(\partial f_0/\partial E)$. 
Because of the $(E-E_{\rm F})$ term, the
numerator may be seen as a centre of mass of 
${\cal T}(E)(\partial f_0/\partial E)$ \cite{CBH11}.
Consequently, both sign and value of $S(T)$ will be sensitive 
even to small changes in the numerator's
integrand below or above $E_{\rm F}$. 

Finally, we note that, in terms of the transmission probability 
${\cal T}(E)$, Mott's formula can be obtained as the 
$T\to 0$ limit of \Eref{SEEBECK} and 
translates into $S(T)$ being
positive (negative) for a negative (positive) 
slope of ${\cal T}(E)$ near the Fermi energy. 
In other words, a large transmission below (above) $E_{\rm F}$
will result in a positive (negative) $S(T)$. Due to this peculiarity,  
the Seebeck coefficient measurement is a well-known
tool in establishing the nature of carriers, $p$- or
$n$-type, in semiconductors.

The above considerations provide the basis to understand
the Seebeck coefficient results, in conjunction with the 
transmission probability profiles depicted in \Fref{NCoCuMUETRANS}(b).
In both magnetic configurations, $S_{\rm P}(T)$ and $S_{\rm AP}(T)$
exhibit a non-monotonous behaviour at low temperatures, consistent
with the changes in the slope of ${\cal T}(E)$ near the Fermi energy
observed in \Fref{NCoCuMUETRANS}(b). These results are consistent with 
Mott's formula. At increased temperatures, both
$S_{\rm P}(T)$ and $S_{\rm AP}(T)$ become and remain positive, with
$S_{\rm AP}(T)$ much larger (about a factor of four) than
$S_{\rm P}(T)$. This positive sign is a direct consequence of the
enhanced transmission in the range of $0.2$~eV below the Fermi energy,
stemming from the minority spin carriers, as discussed above.
It is, as we have seen, the thermoelectric signature of the 
QWS-p-band complex present in the Co slabs of thickness $m=4$~MLs. 

\Fref{NCoCuSeb}(a) further shows that, above $T\simeq 100$~K,
the Seebeck coefficient converges rather fast with the number of
Co repeats $N$, for both magnetic alignments. 
Mathematically, the fast convergence of $S_{\rm P/AP}(T)$
with $N$ results from its definition as a quotient of two integrals. 
The behaviour is clearly different for 
the $T=0$~K conductance $g(T=0)$ which is shown in \Fref{NCoCuSeb}(b)
for the same systems and magnetic configurations. 
Note that, for clarity, we omitted displaying its temperature dependence. 
For both $g_{\rm P}(T)$ and $g_{\rm AP}(T)$ this was 
found to be quite weak, a similar result being reported by 
Kov\'{a}\v{c}ik \etal \cite{KMWB14}.

\begin{figure}[t]
  \centering
  \includegraphics[width=0.48\textwidth]{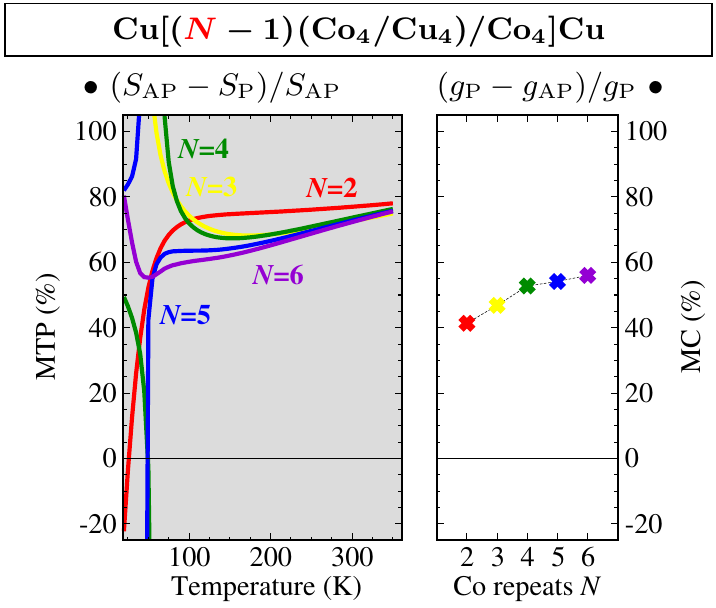}
  \caption{Dependence of the
    magneto-thermopower (MTP/left) and the zero temperature 
    magneto-conductance (MC/right) on the number of Co repeats $N$ 
    in the Cu$[(N-1)$(Co$_4$/Cu$_4$)/Co$_4]$Cu
    multilayered systems.}
  \label{NCoCuMTPMC}
\end{figure}

In order to quantify the magnetic response encountered in the
thermoelectric effect, we calculated the MTP ratio according to
\Eref{MTPrat}. We note that, regardless of the convention 
adopted for the denominator, one problem might always arise 
when plotting a temperature-dependent MTP ratio: Since 
the Seebeck coefficient may change sign as a function
of $T$, one will necessarily encounter discontinuities in
the graphical representation of the MTP ratio.
Such a situation is indeed observed in \Fref{NCoCuMTPMC}, where
we present the calculated MTP and MC ratios for the 
multilayered Cu$[(N-1)$(Co$_4$/Cu$_4$)/Co$_4]$Cu systems.

As can be seen in this figure, very large 
and widely spread values for the MTP
ratio are predicted in the range of low temperatures. These arise in
those areas where $S_{\rm AP}$ approaches zero. Extremely large MTP
ratios have been purposefully omitted from the figure.
Although not as fast as $S_{\rm P}(T)$ and $S_{\rm AP}(T)$, the MTP ratio
also attains convergency with $N$ at high temperatures
and is in general larger than the MC ratio,
a result which is qualitatively consistent with the experimental
findings \cite{BSO00,GFR+04,GSGR+06}. 
As was the case for $g_{\rm P/AP}(T=0)$, the
MC ratio is, in turn, not fully converged with $N$. 
We note that a quantitative comparison with the experimental
data is not attempted here. The reported results
were obtained either for (111)-grown multilayers \cite{BSO00} or
for nanowires of $\sim 10$~nm layer thickness
\cite{GFR+04,GSGR+06}, much larger than 
the $4$~MLs ($6$-$7$~\AA) used in our calculations. 
The effect of the $m$ and $q$  parameters on the 
(magneto)thermoelectric properties of the Co/Cu multilayers
is discussed in the following section.

\subsection{Interlayer exchange coupling}

We close this section by briefly discussing 
the interlayer exchange coupling in the
investigated structures, an electronic structure related issue 
which is closely connected to the GMR and MTP effects.
The functionality of any GMR device relies on its capability
of switching, under an applied magnetic field, 
from an anti-parallel to a parallel
coupling of the magnetisation in the 
adjacent ferromagnetic (FM) layers separated by a non-magnetic (NM) 
spacer. In the absence of an external magnetic field the ground state
magnetic configuration is determined by the
interlayer exchange coupling (IXC). In many FM/NM heterostructures the 
IXC was found to exhibit an oscillatory behaviour with the thickness 
of the NM layer. 

The results we obtained for the IXC in
Cu$[(N-1)$(Co$_m$/Cu$_q$)/Co$_m]$Cu with varying $N$, $m$, and $q$
indicate that an anti-parallel magnetic coupling between
the Co layers is only favoured in the range of thin Cu spacers, 
with an even number, $q=4$ and $q=6$, of MLs. This
appears to be a common feature for all Co layers, irrespective of
their own thickness $m$ and the number of
repeats $N$. Our findings are consistent with previous 
first principles investigations of the IXC in 
Co/Cu bilayers, trilayers or superlattices
\cite{KDTW94,NLZD94,KSK95,LNW+96}. In particular, we note that all
these calculations predict a more stable parallel alignment
for thick Cu spacers in the absence of interface roughness
\cite{LNW+96}.

\section{Multilayers of varying Co and Cu thickness}
\label{CoCuthickvar}

In this section we shall study how thickness changes 
of the individual ferro- and non-magnetic components
(the Co and Cu layers) 
affect the (magneto)thermoelectric properties of the heterostructure.
Particular attention will be given to the QWS-p-band 
hybrid states and their anticipated evolution with the
morphology of the system, as suggested by the 
findings discussed in section~\ref{CoSingleLayer}. 

As demonstrated in \Fref{DOSandTRANSFig}(b), 
for the systems investigated here the
transmission is highly spin-conserving. 
We have also shown that
the QWS-p-hybrids are only present in the minority spin band
and are characterised by a strong localisation at
the Co/Cu interface. An increase of either Co or Cu layer 
thickness is expected to modify the transmission probability profiles
through their influence on the the transmission channels opened
by these states, diminishing their amplitude and weight. In
particular, by removing the large contributions 
to ${\cal T}(E)$ below $E_{\rm F}$, a corresponding change in sign and
increase in absolute value is expected for $S(T)$.

For the results to be presented in the following
we keep a fixed number of repeats $N=4$ in the 
Cu$[(N-1)$(Co$_m$/Cu$_q$)/Co$_m]$Cu system. 
We will start our discussion by focusing
on the changes induced in the transmission 
profiles by the modifications in the thickness $m$ and $q$ of 
the Co and Cu layers and then we will derive the corresponding 
transport properties. The very general expectations formulated above 
will be compared with the results provided by the
actual calculations. We will 
show that the values obtained for both the 
Seebeck coefficient and the MTP span a very broad range, 
depending on the particular $(m,q)$ combination. 
Thus, we conclude that, 
although the expected trends based on 'educated guesses'
are generally fulfilled, in most cases 
explicit calculations are needed in order to make 
accurate predictions \cite{KMWB14,CBH11}.

\subsection{The effect of thicker Co and Cu 
                slabs on the transmission probability profiles}

When discussing the minority-spin BSFs of a single Co layer in
section~\ref{CoSingleLayer} we have emphasised on two important aspects:
(i) the appearance of QWSs and (ii) the existence of an
interface-related, high-mobility p-band. The positions of
the former are obviously thickness dependent but they may 
hybridise with the latter. As a result, strong transmission channels 
for the minority spin carriers are opened.

As shown in \Fref{QWSFig}, this p-band is ever present, 
regardless of the Co thickness. The energy position of the
QWSs, on the other hand, will change as the Co layer becomes thicker:
by a larger extent when $m$ switches from even to odd and
only by a smaller amount for $m\to m+2k$ (identical parity). 
Eventually, the QWSs morph into a continuum
as the thickness of the Co layers further increases. 
This evolution of the QWS-p-band hybrids with $m$ must be accordingly 
reflected in the transmission channels opened by these states. 

\begin{figure*}[t]
  \centering
  \includegraphics[width=0.8666\textwidth]{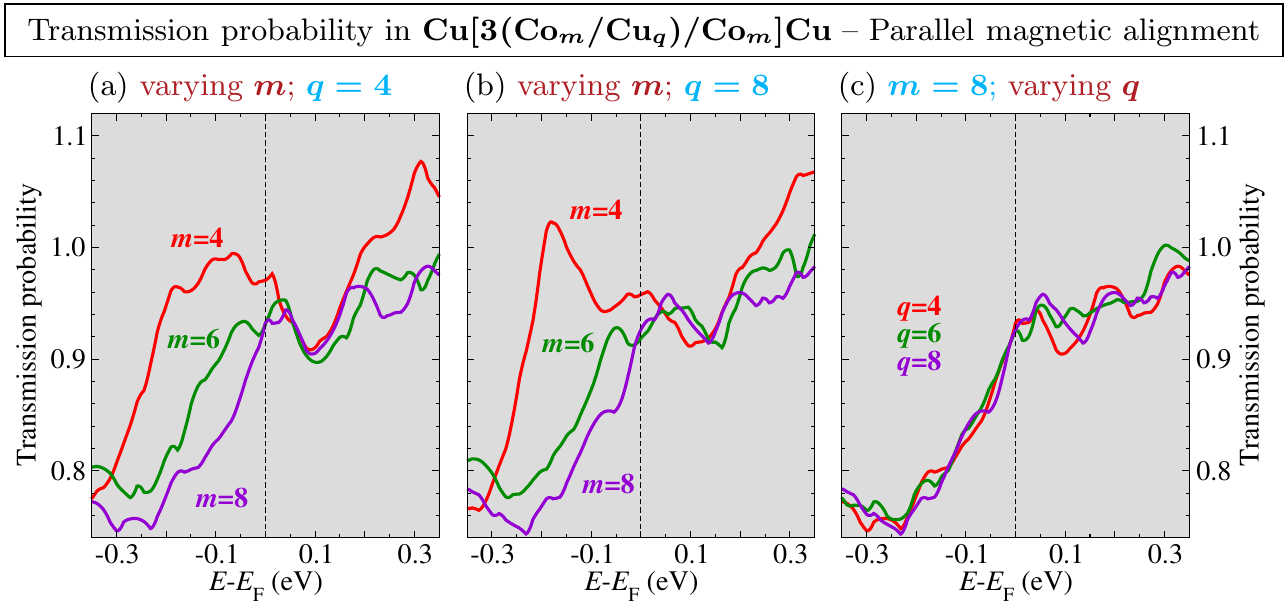}
  \caption{Dependence of the electronic transmission probability
    ${\cal T}(E)$ 
    in Cu$[3$(Co$_m$/Cu$_q$)/Co$_m]$Cu (number of repeats 
    $N=4$) on the thickness
    $m$ and $q$ of the Co and Cu layers: (a) fixed Cu thickness
    $q=4$~MLs, varying Co thickness $m$; (b) fixed Cu thickness
    $q=8$~MLs, varying Co thickness $m$; and (c) fixed
    Co thickness $m=8$~MLs, varying Cu thickness $q$.
    The magnetic configuration corresponds to a
    P-alignment of the Co magnetic moments.
    Transmission profiles for odd number of
    MLs ($m$ and $q$) were skipped for the sake
    of clarity.}
  \label{N4ComCuqTransP}
\end{figure*}

The results displayed in \Fref{N4ComCuqTransP} represent 
a convincing proof that this is indeed the case. Here we show
the calculated transmission profiles for 
Cu$[3$(Co$_m$/Cu$_q$)/Co$_m]$Cu in the parallel (P) magnetic alignment
as a function of either $m$ at fixed $q$ or vice-versa.
For the clarity of the picture, the data for odd number of MLs, 
otherwise in line with the expected trends, have been omitted. 
From left to right, the different 
panels of \Fref{N4ComCuqTransP}
show the changes in ${\cal T}(E)$ 
for (a) $q=4$ and varying $m$, (b) $q=8$ and varying $m$, and
(c) $m=8$ and varying $q$. Note that the varying (fixed)
quantity in the figure is denoted by dark red (light blue)
colours.

Not surprisingly, the thickness dependence of the
QWS-p band complexes discussed above  
has a significant influence on the transmission 
probability profiles. The transmission channels connected
to these states do follow the expected shifts in position 
and intensity. As seen in \Fref{N4ComCuqTransP}(a),
the increase of the Co layers thickness from $m=4$
to $m=6$ and then to $m=8$~MLs causes a dramatic drop in
${\cal T}(E)$ below $E_{\rm F}$. A strong reduction in transmission
can also be observed for the high energy peak at $0.3$~eV. 
The direct comparison of the two panels with constant $q$
values [\Fref{N4ComCuqTransP}(a) with $q=4$~MLs and 
(b) with $q=8$~MLs] makes clear that the smoothening 
of the transmission profiles is mainly caused by the 
variations in the Co thickness, independent of 
the Cu spacer thickness. It is the direct consequence
of the corresponding changes in the electronic structure 
landscape evidenced by the BSFs in \Fref{QWSFig}.

Indeed, one might conclude from \Fref{N4ComCuqTransP}(a) and (b) that
the Cu spacer only plays the role of a 'propagation medium' of varying
size, without too much of an influence on the main features of the
transmission profile. Such an interpretation is apparently 
supported also by the results displayed in \Fref{N4ComCuqTransP}(c) for 
different transmission curves at fixed $m=8$~MLs 
and varying Cu thickness $q$. It is 
only true, however, for thicker Co layers, 
in which case the interface-related effects have a smaller weight.
In the range of thin Co layers (small $m$ values)
the spatial separation of the interfaces
within the {\em non-spin-polarised} spacer
will also affect the transmission profiles, 
albeit in a more subtle way and on a
smaller scale. This can be seen by 
comparing the two curves labelled $m=4$ in the two 
panels (a) and (b) of \Fref{N4ComCuqTransP}.

To summarise, significant changes in the transmission profiles
occur when the thickness of the Co layers is varied. We could
establish a direct connection between these variations and 
the modifications in the electronic structure. In turn, 
a thickness increase of the Cu leads to less spectacular 
changes in ${\cal T}(E)$. Despite the difference in 
the magnitude of the two effects, we will show, 
in the next section, that the Seebeck coefficient as well as 
the MTP are equally sensitive to both $m$ and $q$ variations.

\subsection{Seebeck coefficient and magneto-thermopower
                 for varying Co thickness}

Thermopower measurements on Ni and Fe-Ni
films \cite{APZ12b} have shown that even at a 
$20$~nm thickness of the sample, the Seebeck coefficient 
is about half the value measured for bulk. This is
a general characteristic of nano-structured 
metallic systems and the transition from thin films to bulk 
can be understood as resulting from the reduced weight
of the interface-related transmission channels.
With an increased thickness of the film, the s- and p-states
will have an enhanced contribution to the transmission
above the Fermi energy.
Specific to the currently investigated systems, 
both $S_{\rm P}(T)$ and $S_{\rm AP}(T)$ turn negative
for larger $m$ and $q$ values, with significantly 
increased absolute values.

\begin{figure*}[t]
  \centering
  \includegraphics[width=0.8267\textwidth]{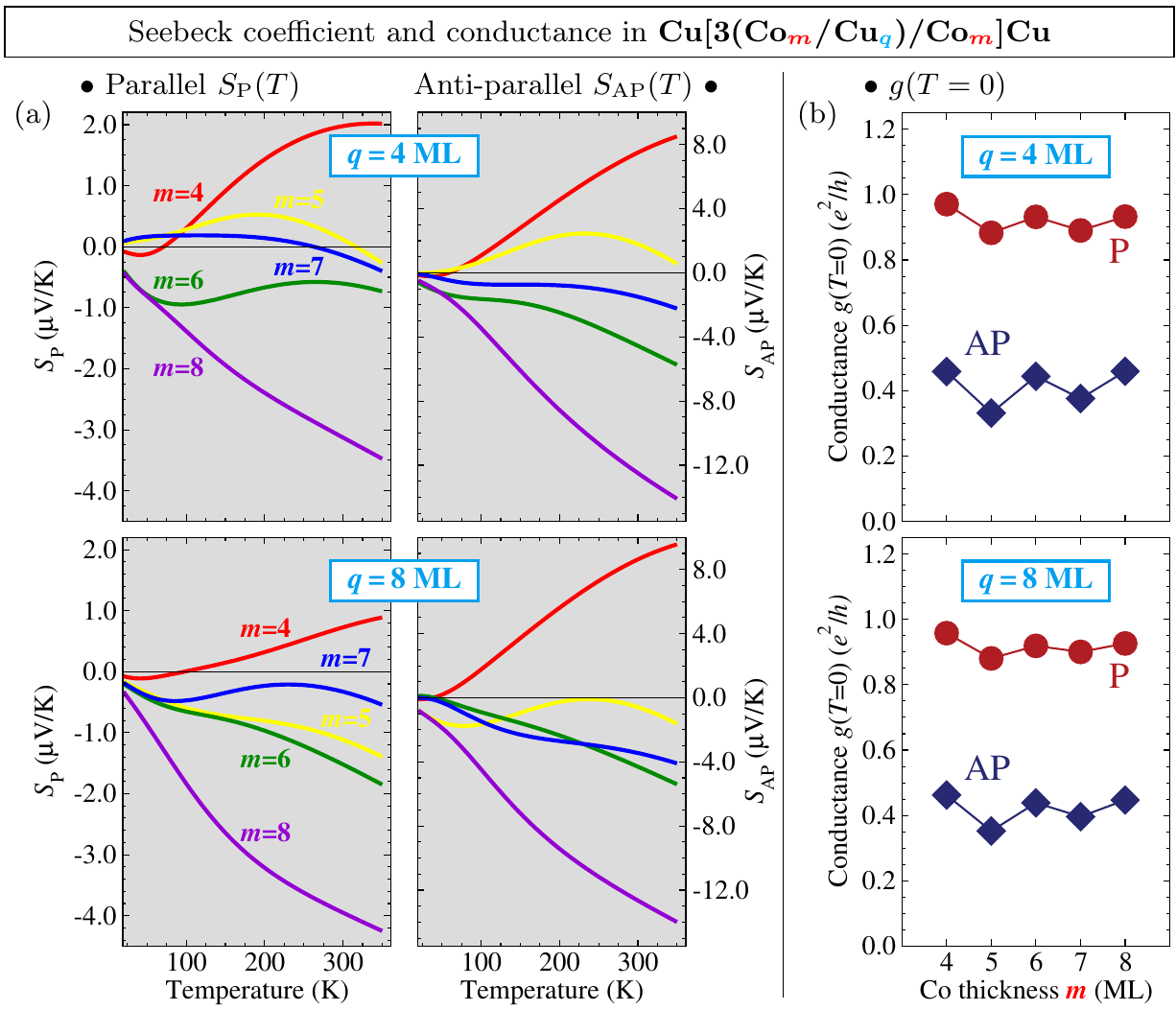}
  \caption{Dependence of (a) the
    Seebeck coefficient $S(T)$ and (b) the zero temperature 
    conductance $g(T=0)$ 
    on the Co layer thickness $m$ (in MLs)
    in the Cu$[3$(Co$_m$/Cu$_q$)/Co$_m]$Cu
    multilayered systems for $q=4$ (top) and $q=8$~MLs (bottom).
    (a) $S_{\rm P}(T)$ for parallel (P, left panel)
    and $S_{\rm AP}(T)$ for anti-parallel (AP, right panel)
    alignment of the magnetisation in the Co layers.
    (b) conductance for P (bullets) and AP (diamonds) alignments.
    Note that, while a different scale is used for $S_{\rm P}(T)$ and 
    $S_{\rm AP}(T)$, each of them remains unchanged when varying Cu
    thickness $q$.}
  \label{N4ComvarCuqSebCond}
\end{figure*}

This behaviour is illustrated in \Fref{N4ComvarCuqSebCond}(a) where we
show the Seebeck coefficients for the parallel
(P, left) and anti-parallel (AP, right) magnetic alignments of
the Cu$[3$(Co$_m$/Cu$_q$)/Co$_m]$Cu multilayered system with a fixed
Cu spacer thickness $q=4$~MLs (top) and $q=8$~MLs (bottom),
for varying Co thickness $m$. 
The most spectacular result, anticipated from the changes in the
transmission profiles, is the change in sign obtained for both
$S_{\rm P}(T)$ and $S_{\rm AP}(T)$ when $m$ increases. 
Note that the same colour-coding convention (from red to
purple for increasing $m$) is used in this figure
as introduced above. The 
$m$-dependence of $S(T)$ patterns, both in P- and AP-alignment, 
show remarkable similarities for the two $q$ values, indicating the
less important role played by the Cu spacer in governing the
(magneto)thermoelectric properties of the investigated systems.
Notable differences can only be seen for 
$m=5$ and $m=7$~MLs Co. In these systems
the Seebeck coefficient has a small absolute value 
and fluctuating sign, as seen, for example in
the $S_{\rm P}(T)$ corresponding to $(m,q)=(5,4)$ and $(5,8)$.

For the same systems and geometrical parameters we show
in \Fref{N4ComvarCuqSebCond}(b) the calculated zero-temperature 
conductance $g(T=0)$. In contrast to the Seebeck coefficient, 
the conductance is seen to exhibit much less fluctuations 
with $m$ and $q$. The reason for this
different behaviour lies, once again, in the actual energy range
where the transmission probability is changing with $m$. As we could
see, this essentially takes place $\pm 0.2$~eV away from the Fermi
energy and, as such, is not affecting $g(T=0)$. 
Quantum confinement effects do manifest, however, 
also in the conductance: A clear separation in 
the $m$ dependence for even and odd values is evidenced in 
\Fref{N4ComvarCuqSebCond}(b). This originates from the
parity dependence of the standing waves formed inside the Co 
layers by the interaction of the interface states at either sides.

\begin{figure*}[t]
  \centering
  \includegraphics[width=0.9333\textwidth]{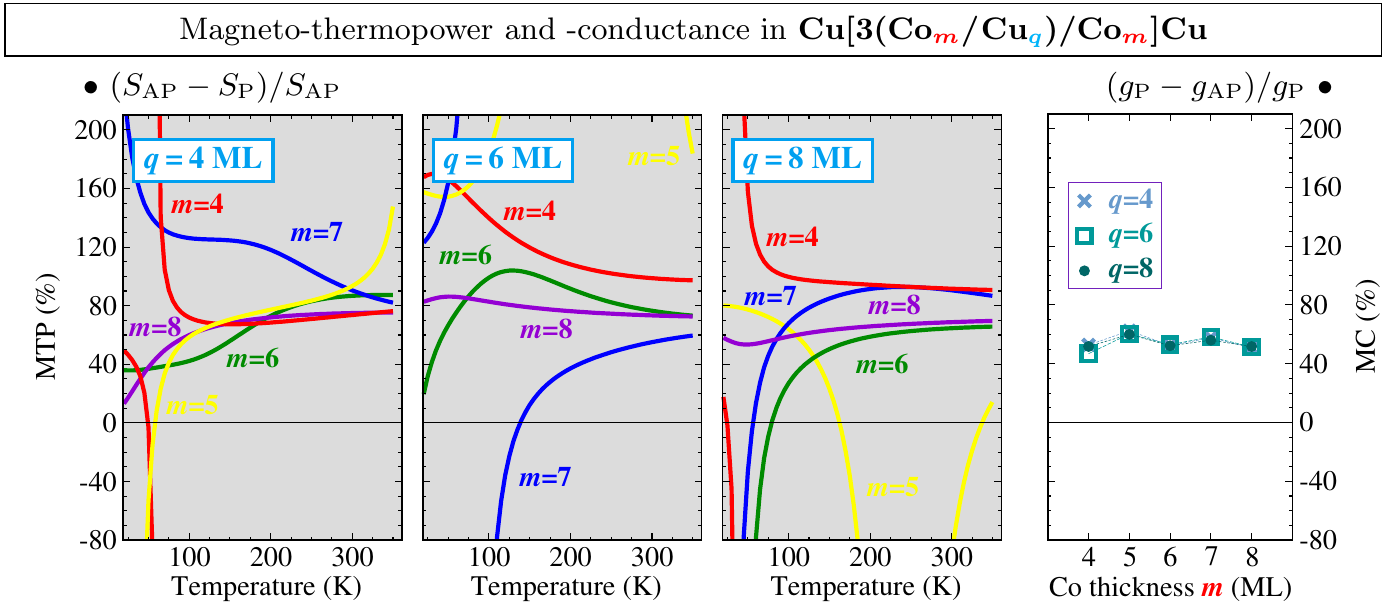}
  \caption{
    Dependence of the
    magneto-thermopower (MTP/left) and the zero temperature 
    magneto-conductance (MC/right)
    on the Co layer thickness $m$ (in MLs) 
    in the Cu$[3$(Co$_m$/Cu$_4$)/Co$_m]$Cu
    multilayered systems.}
  \label{N4ComCu468MTPMC}
\end{figure*}

Coming back to the Seebeck coefficient calculated for the
two magnetic alignments, P and AP, we note that, 
regardless of the explicit $m$ and $q$ values, large
differences are predicted between 
$S_{\rm AP}(T)$ and $S_{\rm P}(T)$. Since, on the other hand,
these differences are not independent of $T$, a rather 
broad range of values can be expected for the MTP. 
The temperature dependent MTP ratios are shown in 
\Fref{N4ComCu468MTPMC} (left panel) for various Cu spacer 
thickness $q$ and compared with the zero 
temperature MC (right panel) on an identical scale.

As can be seen in this figure, the range of attained 
MTP values is much broader than that of the corresponding MC.
The latter exhibits slight fluctuations with $m$, but it remains
in an interval of $40$-$60$~\%, rather independent of the
Cu spacer thickness $q$. Note that, as was the case for 
the $N$-varying systems, exceedingly large values of the
MTP ratios, caused by $S_{\rm P}(T)$ approaching zero, are not
displayed. For temperatures higher than $100$~K, the MTP
is obviously larger than the MC, essentially
any MTP ratio between $40$ and $100$~\% being accessible
by an appropriate ($m$, $q$) selection. 

While this is merely of theoretical interest,
as not any ($m$,$q$)-combination is necessarily 
attainable experimentally or energetically stable, 
the results point out to an important aspect.
As far as the electronic structure contribution to the transport
properties is concerned, using a thermal gradient rather 
than an electric field could indeed be more advantageous 
in order to gain a large magnetic sensitivity in a 
magnetic read-out device.

\subsection{Seebeck coefficient and magneto-thermopower
               for varying Cu thickness}

In order to complete our discussion, we analyse how the change
on the Cu spacer size, at a fixed Co thickness, 
influences the transport properties of the 
Cu$[3$(Co$_m$/Cu$_q$)/Co$_m]$Cu multilayered
systems. As illustrated by the transmission profiles shown 
in \Fref{N4ComCuqTransP}(c)
for fixed $m=8$~MLs and varying $q$, 
the role of the Cu thickness in modelling the 
transmission is minor: 
Once a basic shape in the energy dependence of the 
transmission is set by the given Co thickness $m$, i.e., by filtering
and smoothening the QWS-related channels, no significant changes
occur in ${\cal T}(E)$ as $q$ increases. 

\begin{figure*}[t]
  \centering
  \includegraphics[width=0.8267\textwidth]{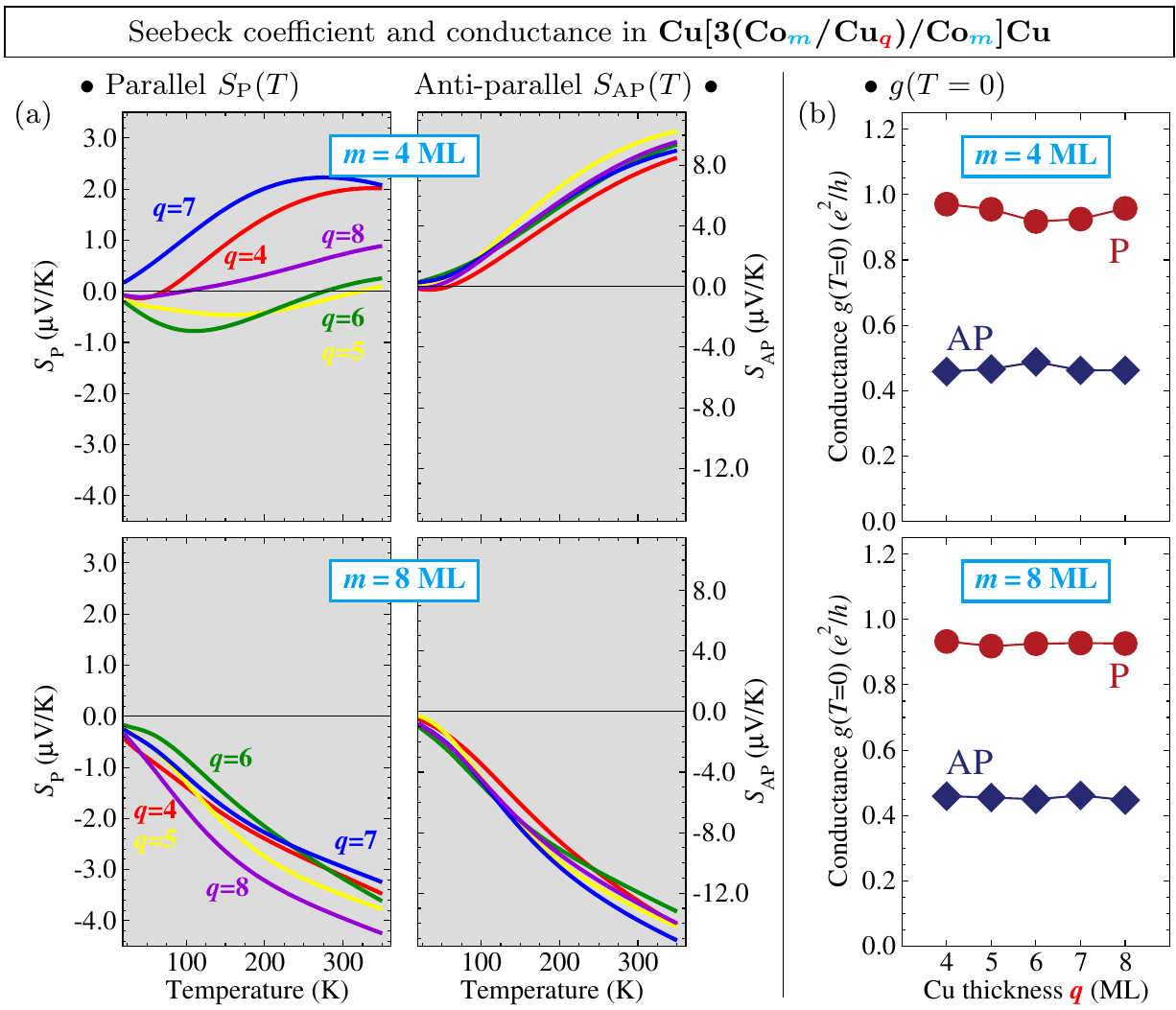}
  \caption{Same as \protect\Fref{N4ComvarCuqSebCond} but
    showing the variation of $S_{\rm P}(T)$, $S_{\rm AP}(T)$,
    and $g(T=0)$ in Cu$[3$(Co$_m$/Cu$_q$)/Co$_m]$Cu
    with the Cu layer thickness
    $q$ for $m=4$ (top) and $m=8$ (bottom).
    Note that in panel (a) the same scale is used column-wise, that is,
    when comparing $S_{\rm P/AP}(T)$ for varying Co thickness $m$, while
    the scales are different for $S_{\rm P}(T)$ and $S_{\rm AP}(T)$ at
    equal $m$.}
  \label{N4ComCuqvarSebCond}
\end{figure*}

This is accordingly reflected in the temperature dependence of 
the Seebeck coefficient for both P- and AP-alignments
shown in \Fref{N4ComCuqvarSebCond}(a). The $S_{\rm P}(T)$ and 
$S_{\rm AP}(T)$ curves corresponding to various $q$ values are
grouped together over the whole temperature range. 
The notable exception is the $m=4$~MLs system in P-alignment, 
where the formed QWS-p-band complexes  
couple stronger across the Cu spacer.
In contrast, for $m=4$~MLs in the AP-alignment 
the $S_{\rm AP}(T)$ values of the different spacer thickness
are fairly close to another because the electron scattering is
essentially spin-conserving. For the AP magnetic configuration 
this corresponds to an effective spacer between 
the Co layers larger than the actual, physical one. 
For a Co thickness $m=8$~MLs, equivalent to an absence 
of the QWS-p-band enabled channels, the variations 
with $q$ of both $S_{\rm P}(T)$ and $S_{\rm AP}(T)$ is 
even more reduced. This behaviour is consistent with the 
experimental findings of Shi \etal~\cite{SYPS93} in Co/Cu multilayers 
of comparable thickness but in a CIP geometry.

The corresponding zero temperature conductance results 
shown in \Fref{N4ComCuqvarSebCond}(b) follow a similar characteristic
of a rather weak dependence on $q$ at a fixed $m$. 
We note in particular the complete absence of any oscillations 
in $g(T=0)$ between odd and even $q$, as was the case of varying $m$.

\begin{figure}[t]
  \centering
  \includegraphics[width=0.48\textwidth]{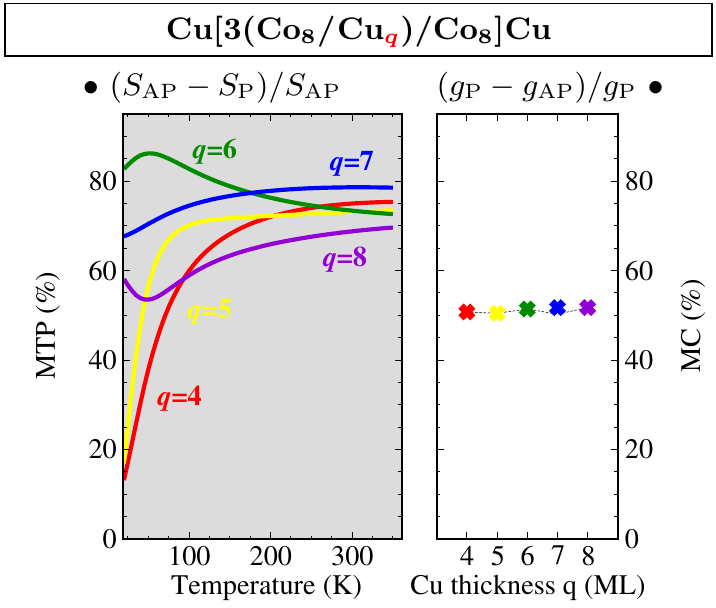}
  \caption{Dependence of the
    MTP (left panel) and the zero temperature 
    MC (right panel) on the number of Cu
     layer thickness $q$ in the 
    Cu$[3$(Co$_m$/Cu$_q$)/Co$_m]$Cu multilayered systems.}
  \label{N4Co8CuqMTPMC}
\end{figure}

In spite of the much smaller spread over varying $q$
of both $S_{\rm P}(T)$ and $S_{\rm AP}(T)$ at fixed $m=8$~MLs, 
the MTP ratio still exhibits a broad range of values,
as shown in the left panel of \Fref{N4Co8CuqMTPMC}.
Above $\simeq 100$~K, however, the temperature dependence 
of the MTP for a fixed $q$ is much weaker, even for small
values of the Cu spacer thickness. In contrast to the MTP,
the MC ratio (right panel) is almost independent of $q$,
showing small fluctuations around $50$~\%. This proves, once again,
that the MTP offers in principle a much larger sensitivity 
to small changes in the electronic structure than the MC. It also
implies that the reproducibility of independent experimental
data might turn into a problematic issue. In addition,
we note that the analysis of the results of this section, 
encompassing different $(m,q)$ combinations, does not lead to
any obvious correlation between the MC and the MTP, 
nor does enable us to make a definite statements about 
one configuration being better suited than another for an enhanced MTP. 

\section{Conclusions}

In summary, we have presented results of {\em ab initio} 
calculations of the magneto-thermoelectric properties 
for a series of Co/Cu multilayered systems embedded in Cu(001) 
with the general formula Cu$[(N-1)$(Co$_m$/Cu$_q$)/Co$_m]$Cu. 
Our investigations focused on the influence the various
morphological parameters | number of repeats $N$, 
layer thickness $m$ and $q$ |
have upon the underlying
electronic structure and, through the induced modifications, 
on the various transport properties of the heterostructures. 

While adopting a spin-polarised fully relativistic formalism,
we have nevertheless found that the electronic transmission 
in the Co/Cu multilayers is to a large extent spin-conserving. 
For thin Co layers ($m\leq 7$~MLs) the minority spin 
channel, although weaker than the majority spin one, 
strongly modulates the transmission profile. This modulation
is caused by quantum well states present in the Co layer which 
hybridise with a Co/Cu interface-related p-band, 
opening this way very efficient transmission channels.
We have shown that the energy position of
the such formed hybrid states, which varies as a function
of the Co layer thickness, is accordingly reflected
in the transmission profiles. Significant changes with 
$m$ occur, leading to a large sensitivity of the Seebeck 
coefficient and the magnetothermopower (MTP) to the thickness 
of the Co layers. The other geometrical parameters, $N$ and $q$, have
a much smaller influence on the transport properties.
We need to emphasise, however, on the broad range of values
that both the Seebeck coefficient and the MTP may attain in
such systems, depending on their morphology, which might
cause difficulties when comparing theoretical results with
experimental ones. Further effects, not considered here, such as
thickness fluctuations, interface interdiffusion or relaxation,
defect formation in grown multilayered components, 
might increase the spread of the results even more.

A comparison of the MTP behaviour to that of the 
magneto-conductance at the same geometry leads to the conclusion 
that an MTP read-out of the magnetisation state 
can be equally or even more efficient than a GMR-based device.
Our results suggest that a small number of Co layers with 
precise control of the layer thickness
may be more advantageous for this purpose than 
increasing the number of the Co repeats. 

\ack

This work was supported by the German Research Foundation
({\em Deutsche Forschungsgemeinschaft -- DFG}) 
within the Priority Program 1538 ''Spin Caloric Transport
(SpinCaT)''. 
The authors gratefully acknowledge the computing time granted by the 
John von Neumann Institute for Computing (NIC) and provided on the 
supercomputer JUROPA at J\"ulich Supercomputing Centre (JSC).
Additional computer facilities have been offered by the
Center for Computational Sciences and Simulation
(CCSS) at the University Duisburg-Essen. 

\section*{References}

\begin{thebibliography}{10}

\bibitem{GB97}
Gijs M~A~M and Bauer G~E~W, 1997, {\em Advances in Physics}, 46:285--445.

\bibitem{CBM+91}
Conover M~J, Brodsky M~B, Mattson J~E, Sowers C~H, and Bader S~D, 1991, {\em
  J.\ Magn.\ Magn.\ Mat.}, 102:L5--L8.

\bibitem{PFS+92}
Piraux L, Fert A, Schroeder P, Loloee R, and Etienne P, 1992, {\em J.\ Magn.\
  Magn.\ Mat.}, 110:L247--L253.

\bibitem{SYPS93}
Shi J, Yu R~C, Parkin S~S~P, and Salamon M~B, 1993, {\em Journal of Applied
  Physics}, 73:5524--5526.

\bibitem{NSH+94}
Nishimura K, Sakurai J, Hasegawa K, Saito Y, Inomata K, and Shinjo T, 1994,
  {\em Journal of the Physical Society of Japan}, 63:2685--2690.

\bibitem{BSO00}
Baily S~A, Salamon M~B, and Oepts W, 2000, {\em Journal of Applied Physics},
  87:4855--4857.

\bibitem{SMAZF12}
Scharf B, Matos-Abiague A, \v{Z}uti\'{c} I, and Fabian J, Feb 2012, {\em Phys.
  Rev. B}, 85:085208.

\bibitem{GFR+04}
Gravier L, F\'{a}bi\'{a}n A, Rudolf A, Cachin A, Wegrowe J~E, and Ansermet J~P,
  2004, {\em J.\ Magn.\ Magn.\ Mat.}, 271:153--158.

\bibitem{GSGR+06}
Gravier L, Serrano-Guisan S, Reuse F, and Ansermet J~P, 2006, {\em Phys.\ Rev.\
  B}, 73:024419.

\bibitem{BVM+13}
B\"ohnert T, Vega V, Michel A~K, Prida V~M, and Nielsch K, 2013, {\em Applied
  Physics Letters}, 103:092407.

\bibitem{BNM+14}
B\"ohnert T, Niemann A~C, Michel A~K, B\"a\ss{}ler S, Gooth J, T\'oth B~G,
  Neur\'ohr K, P\'eter L, Bakonyi I, Vega V, Prida V~M, and Nielsch K, 2014,
  {\em Phys. Rev. B}, 90:165416.

\bibitem{HKL+14}
Hu X~K, Krzysteczko P, Liebing N, Serrano-Guisan S, Rott K, Reiss G, Kimling J,
  B\"ohnert T, Nielsch K, and Schumacher H~W, 2014, {\em Appl. Physics Lett.},
  104:092411.

\bibitem{MJ58}
Mott N~F and Jones H.
\newblock {\em The Theory of the Properties of Metals and Alloys}.
\newblock Dover Publications, 1958.

\bibitem{KDTW94}
Kudrnovsk\'y J, Drchal V, Turek I, and Weinberger P, 1994, {\em Phys. Rev. B},
  50:16105--16108.

\bibitem{NLZD94}
Nordstr\"om L, Lang P, Zeller R, and Dederichs P~H, 1994, {\em Phys. Rev. B},
  50:13058--13061.

\bibitem{KSK95}
Krompiewski S, S\"uss F, and Krey U, 1995, {\em J.\ Magn.\ Magn.\ Mat.},
  149:L251--L254.

\bibitem{LNW+96}
Lang P, Nordstr\"om L, Wildberger K, Zeller R, Dederichs P~H, and Hoshino T,
  1996, {\em Phys. Rev. B}, 53:9092--9107.

\bibitem{BZN+96}
Butler W~H, Zhang X~G, Nicholson D~M~C, Schulthess T~C, and MacLaren J~M, 1996,
  {\em J.\ Appl.\ Phys.}, 79:5282.

\bibitem{WBA+96}
Weber W, Bischof A, Allenspach R, W\"ursch C, Back C~H, and Pescia D, 1996,
  {\em Phys. Rev. Lett.}, 76:3424--3427.

\bibitem{SUB+97}
Szunyogh L, \'Ujfalussy B, Blaas C, Pustogowa U, Sommers C, and Weinberger P,
  1997, {\em Phys.\ Rev.\ B}, 56:14036.

\bibitem{XZT+06}
Xia K, Zwierzycki M, Talanana M, Kelly P~J, and Bauer G~E~W, 2006, {\em Phys.
  Rev. B}, 73:064420.

\bibitem{BZM00}
Binder J, Zahn P, and Mertig I, 2000, {\em J.\ Appl.\ Phys.}, 87:5182.

\bibitem{BSW+02}
Blaas C, Szunyogh L, Weinberger P, Sommers C, Levy P~M, and Shi J, 2002, {\em
  Phys. Rev. B}, 65:134427.

\bibitem{SW05}
Sommers C and Weinberger P, 2005, {\em Phys. Rev. B}, 72:054431.

\bibitem{PK13}
Popescu V and Kratzer P, 2013, {\em Phys. Rev. B}, 88:104425.

\bibitem{KMWB14}
Kov\'a\v{c}ik R, Mavropoulos P, Wortmann D, and Bl\"ugel S, 2014, {\em Phys.
  Rev. B}, 89:134417.

\bibitem{PEP+04}
Popescu V, Ebert H, Papanikolaou N, Zeller R, and Dederichs P~H, 2004, {\em
  Journal of Physics: Condensed Matter}, 16:S5579.

\bibitem{PEP+05}
Popescu V, Ebert H, Papanikolaou N, Zeller R, and Dederichs P~H, 2005, {\em
  Phys. Rev. B}, 72:184427.

\bibitem{EKM11}
Ebert H, K{\"o}dderitzsch D, and Min\'{a}r J, 2011, {\em Rep. Prog. Phys.},
  74:096501.

\bibitem{SPR-TB-KKR}
{\em The Munich SPR-TB-KKR package}, \mbox{H.~Ebert~{\em et~al.}} \newline
  http://www.ebert.cup.uni-muenchen.de/spr-tb-kkr.

\bibitem{SUWK94}
Szunyogh L, \'Ujfalussy B, Weinberger P, and Koll\'ar J, 1994, {\em Phys. Rev.
  B}, 49:2721--2729.

\bibitem{ZDU+95}
Zeller R, Dederichs P~H, \'{U}jfalussy B, Szunyogh L, and Weinberger P, 1995,
  {\em Phys.\ Rev.\ B}, 52:8807.

\bibitem{WZD97}
Wildberger K, Zeller R, and Dederichs P~H, 1997, {\em Phys. Rev. B},
  55:10074--10080.

\bibitem{BS89}
Baranger H~U and Stone A~D, 1989, {\em Phys.\ Rev.\ B}, 40:8169.

\bibitem{MPD04}
Mavropoulos P, Papanikolaou N, and Dederichs P~H, 2004, {\em Phys.\ Rev.\ B},
  69(15):125104.

\bibitem{VWN80}
Vosko S~H, Wilk L, and Nusair M, 1980, {\em Canadian Journal of Physics},
  58:1200.

\bibitem{TDK+97}
Turek I, Drchal V, Kudrnovsk\'y J, \v{S}ob M, and Weinberger P.
\newblock {\em Electronic Structure of Disordered Alloys, Surfaces and
  Interfaces}.
\newblock Kluwer Academic Publishers, Boston, 1997.

\bibitem{SI86}
Sivan U and Imry Y, 1986, {\em Phys. Rev. B}, 33:551--558.

\bibitem{KCME13}
K\"odderitzsch D, Chadova K, Min\'ar J, and Ebert H, 2013, {\em New Journal of
  Physics}, 15(5):053009.

\bibitem{WSvsB09}
Wysocki A~L, Sabirianov R~F, van Schilfgaarde M, and Belashchenko K~D, 2009,
  {\em Phys. Rev. B}, 80:224423.

\bibitem{KDT+12}
Kudrnovsk\'y J, Drchal V, Turek I, Khmelevskyi S, Glasbrenner J~K, and
  Belashchenko K~D, 2012, {\em Phys. Rev. B}, 86:144423.

\bibitem{MN14}
Mahfouzi F and Nikoli\'{c} B~K, 2014, {\em Phys. Rev. B}, 90:045115.

\bibitem{IK06}
Irkhin V~Y and Katsnelson M~I, 2006, {\em Phys. Rev. B}, 73:104429.

\bibitem{CSA+08}
Chioncel L, Sakuraba Y, Arrigoni E, Katsnelson M~I, Oogane M, Ando Y, Miyazaki
  T, Burzo E, and Lichtenstein A~I, 2008, {\em Phys. Rev. Lett.}, 100:086402.

\bibitem{JM80}
Jonson M and Mahan G~D, 1980, {\em Phys. Rev. B}, 21:4223--4229.

\bibitem{CBH11}
Czerner M, Bachmann M, and Heiliger C, 2011, {\em Phys. Rev. B}, 83:132405.

\bibitem{APZ12b}
Avery A~D, Pufall M~R, and Zink B~L, 2012, {\em Phys. Rev. B}, 86:184408.

\end{thebibliography}

\end{document}